\documentclass[titlepage,12pt]{article}
\usepackage[dvipdfmx]{graphicx}
\usepackage{textcomp}
\usepackage{amssymb,amsmath,color,graphics,amscd,epsf,indentfirst,amsfonts}
\usepackage{mathtools}
\usepackage{enumerate}

\font\mybb=msbm10 at 12pt
\def\bb#1{\hbox{\mybb#1}}
\def\Z {\bb{Z}}
\def\R {\bb{R}}

\font\mycc=msbm10 at 10pt
\def\cc#1{\hbox{\mycc#1}}
\def\sZ {\cc{Z}}

\def\cD {{\cal D}}
\def\cE {{\cal E}}
\def\cH {{\cal H}}

\def\cN {{\cal N}}
\def\cO {{\cal O}}

\def\diag{\mathop{\rm diag}\nolimits}

\def\tr{\mathop{\rm tr}\nolimits}
\def\mod{\mathop{\rm mod}\nolimits}

\def\mat#1{\matt[#1]}
\def\matt[#1,#2,#3,#4]{\left(%
\begin{array}{cc} #1 & #2 \\ #3 & #4 \end{array} \right)}

\def\ethe[#1,#2]{\vartheta\left[#1\atop #2\right]}

\makeatletter
\@addtoreset{equation}{section}
\@addtoreset{footnote}{page}
\makeatother

\newcommand{\wt}{\widetilde}
\newcommand{\wh}{\widehat}
\newcommand{\ol}{\overline}

\newcommand{\ra}{\rightarrow}

\newcommand{\nn}{\nonumber}

\newcommand{\half}{\frac{1}{2}\,}
\newcommand{\VEV}[1]{\left\langle #1\right\rangle}
\newcommand{\bra}[1]{\left\langle\, #1\,\right|}
\newcommand{\ket}[1]{\left|\, #1\,\right\rangle}

\newcommand{\del}{\partial}

\newcommand{\AD}[1]{$\ol{\mbox{D~\,}}\!\!\!#1$}
\newcommand{\Op}[1]{$\mbox{O}#1^+$}
\newcommand{\Opm}[1]{$\mbox{O}#1^\pm$}
\newcommand{\Om}[1]{$\mbox{O}#1^-$}
\newcommand{\Omt}[1]{$\wt{\mbox{O}#1}^-$}

\newcommand{\oneform}{{\rm 1\mathchar`-form}}
\newcommand{\ax}{{\rm axial}}

\def\beq{\begin{equation}}
\def\eeq{\end{equation}}

\numberwithin{equation}{section}

\begin{document}
\begin{titlepage} 
\begin{flushright}
\hfill{YITP-18-133}\\
\end{flushright}
\vskip .7cm
\begin{center}
\font\titlerm=cmr10 scaled\magstep4
    \font\titlei=cmmi10 scaled\magstep4
    \font\titleis=cmmi7 scaled\magstep4
    \centerline{\LARGE \titlerm 
      Vacuum Structure of Charge $k$ Two-Dimensional QED}
      \vskip 0.3cm
    \centerline{\LARGE \titlerm
      and Dynamics of an Anti D-String Near an O$1^-$-plane}
    \vskip 0.3cm
\vskip 1cm
{Adi Armoni$^\star$ and Shigeki Sugimoto$^{\natural,\sharp}$}\\
\vskip 0.5cm
       {\it $^\star$Department of Physics, College of Science}\\
       {\it Swansea University, SA2 8PP, UK}\\
\medskip
{\it $^\natural$Center for Gravitational Physics, Yukawa Institute for
Theoretical Physics,\\
Kyoto University, Kyoto 606-8502, JAPAN}\\
{\it {$^{\sharp}$Kavli Institute for the Physics and Mathematics
 of the Universe (WPI),\\
The University of Tokyo, Kashiwanoha, Kashiwa 277-8583, JAPAN
}}\\ 
\medskip
\vskip 0.5cm
{$^\star$a.armoni@swansea.ac.uk, $^\natural$sugimoto@yukawa.kyoto-u.ac.jp}\\

\end{center}
\vskip .5cm
\centerline{\bf Abstract}

\baselineskip 20pt
%

\vskip .5cm 
\noindent

We study the vacuum structure of $N_f$ flavour two-dimensional QED
with an arbitrary integer charge $k$. We find that the axial symmetry
is spontaneously broken from $\Z_{kN_f}$ to $\Z_{N_f}$ due to
the non-vanishing condensate of a flavour singlet operator,
resulting in $k$ degenerate vacua.
An explicit construction of the $k$ vacua is given by using
a non-commutative algebra obtained as a central extension
of the $\Z_{kN_f}$ discrete axial symmetry and $\Z_{k}$ 1-form
(center) symmetry, which represents the mixed 't Hooft anomaly between
them.

We then give a string theory realization of such a system
with $k=2$ and $N_f=8$ by putting an anti D-string in the vicinity of
an orientifold O$1^-$-plane and study its dynamics using the
two-dimensional gauge theory realized on it. We calculate the potential
between the anti D-string and the O$1^-$-plane and find repulsion in
both weak and strong coupling regimes of the two-dimensional gauge
theory, corresponding to long and short distances, respectively. We also
calculate the potential for the $(Q,-1)$-string (the bound state of an
anti D-string and $Q$ fundamental strings) located close to the
O$1^-$-plane.
The result is non-perturbative in the string coupling.
 
\vfill
\noindent
\end{titlepage}\vfill\eject

\setcounter{equation}{0}

\pagestyle{empty}
\small
\vspace*{-0.7cm}

\normalsize
\pagestyle{plain}
\setcounter{page}{1}

\section{Introduction} 
\label{intro}

The study of the dynamics of string theory at strong coupling is
hard. Both string perturbation theory and non-perturbative string
phenomenon are difficult to calculate. Field theory calculations are
often easier especially in the case of two-dimensional gauge theories.

The purpose of this paper is twofold. In its first part we study a
variant of two-dimensional QED (from now on abbreviated ``2 dim QED'')
with $N_f$ flavours of Dirac fermion fields of
charge $k$, where $k$ is an arbitrary integer. (In the following,
we assume $k$ to be positive without loss of generality.)
This is a generalization of the well-known multi flavour Schwinger
model.\footnote{The $N_f=1$ case is recently analyzed in
\cite{Anber:2018jdf}.}
Although the $k$ dependence in the action of 2 dim QED can be
eliminated by rescaling the gauge field and the gauge coupling,
it enters in the flux quantization condition and
the charge $k$ is actually physically relevant.
In fact, the $k$ dependence appears in the symmetry of the system.
When the fermions are massless,
it has $\Z_{kN_f}$ anomaly free discrete axial symmetry and $\Z_k$
1-form (center) symmetry, which play crucial roles in our analysis.
We argue that the $\Z_{kN_f}$ discrete axial symmetry is spontaneously
broken to $\Z_{N_f}$ and, as a result, there are $k$
distinct vacua, generalizing the result for $N_f=1$ given in
\cite{Anber:2018jdf}.
Interestingly, when $N_f>1$, the chiral condensate
$\VEV{\ol\psi_i\psi^j}$ vanishes,
because the chiral $SU(N_f)_L\times SU(N_f)_R$ symmetry
cannot be spontaneously broken due to the Coleman-Mermin-Wagner
theorem \cite{Coleman:1973ci,Mermin:1966fe,Hohenberg:1967zz},
but nonetheless the axial symmetry is spontaneously broken
because the vacuum expectation value of the determinant
of the fermion bilinear operator $\det(\psi_{Rj}^\dag\psi_L^i)$ is
non-vanishing.\footnote{
This scenario was also suggested in \cite{Anber:2018xek}.
See \cite{Unsal:2007jx,Anber:2018tcj,Yamaguchi:2018xse}
for analogous phenomena in 4 dim gauge theory.}
The spontaneous break down of the axial symmetry is also
characterized by the mixed 't Hooft anomaly between
the $\Z_{kN_f}$ axial symmetry and the $\Z_k$ 1-form symmetry
as discussed in \cite{Anber:2018jdf} for $N_f=1$ case.
The existence of the mixed 't Hooft anomaly is understood
as the fact that the axial $\Z_{kN_f}$ symmetry and
the $\Z_k$ 1-form symmetry are centrally extended
in the quantum theory.\cite{Gaiotto:2017yup}
The centrally extended algebra is non-commutative
and gives a stringent constraint on the vacuum structure.
We give an explicit construction of the $k$ vacua
by utilizing this non-commutative algebra.

We mainly work in a bosonized description of 2 dim QED.
We check that all the global symmetry as well as
the mixed 't Hooft anomaly are realized in the bosonized
description and the results for $N_f=1$ given
in \cite{Anber:2018jdf} are reproduced in a simplified way.
It turns out to be very efficient in the generalization to $N_f>1$.
We also discuss how the vacuum degeneracy is lifted when a small mass
(with respect to the gauge coupling) is given to the fermion.

In the second part of the paper we use the results obtained in the first
part to study the dynamics of a non-supersymmetric brane
configuration. We will focus on a system that consists of an orientifold
\Om1-plane and a \AD1-brane. Here, the \AD1-brane is the anti D-string
obtained by flipping the orientation of the D1-brane
in the maximally supersymmetric \Om1-D1 system.
The O$p$-\AD p system is a non-supersymmetric system
with no tree level tachyon fields.
It has been studied as a mechanism to break supersymmetry
in a controlled way
\cite{Sugimoto:1999tx,Antoniadis:1999xk,Aldazabal:1999jr,Angelantonj:1999ms}
and provides interesting playgrounds
to study non-supersymmetric quantum field theories
\cite{Uranga:1999ib,Sugimoto:2012rt,Armoni:2013ika,Armoni:2014cia,Armoni:2017jkl}.
We show that the \Om1-\AD1 system contains a sector that is
described by 2 dim QED with $k=2$ and $N_f=8$.

As an application of the analysis in the first part, we calculate the
potential between the \Om1-plane and the \AD1-brane. The distance
between them is parametrized by the value of scalar fields that
correspond to fermion mass in 2 dim QED.
Using the standard relations between gauge theory and string theory
parameters, the potential is calculated by evaluating
the vacuum expectation value of the Hamiltonian of 2 dim QED
as a function of the fermion mass.
Since the gauge theory is super-renormalizable, the short distance
potential is controlled by strong coupling and the long distance
potential is controlled by weak coupling dynamics. 
At short distances the potential is calculated by
using the strong coupling analysis in 2 dim QED. The result contains
fractional powers of string coupling $g_s$, which clearly shows
that the result is non-perturbative.
In the weak coupling regime, we use the 1-loop Coleman-Weinberg
potential to calculate the potential. 

We find that the \AD1-brane is repelled from the
\Om1-plane at both short and long distance, which suggests that there is
a run away potential. We also calculate the potential for
\AD1-brane with $Q$ unit of electric flux on it, which corresponds
to the bound state of the \AD1-brane and $Q$ fundamental strings.

The paper is divided into two main parts and it is organized as follows:
in section \ref{vacuum}, we discuss the vacuum structure of the
charge $k$ multi flavour 2 dim QED.
In section \ref{AppString}, we discuss the \Om1-\AD1 system and its
dynamics. Section \ref{conclusions} is devoted to an
outlook and a discussion about future directions.

\section{Vacuum Structure of Charge $k$ Multi Flavour QED} 
\label{vacuum}

In this section, we wish to find the vacuum structure of 2 dim QED
with massless $N_f$ flavours of charge $k$ fermions.
The model is exactly solvable using bosonization. We also consider
the massive case in section \ref{MassDeform}.

\subsection{Symmetry and anomaly}
\label{symanom}

The action of the system we consider is given by
\begin{eqnarray}
 S_{\rm QED}=\int d^2x \left(
-\frac{1}{4 e^2}F_{\mu\nu}^2+i\ol\psi_i\gamma^\mu\left(
\del_\mu+ikA_\mu
\right)\psi^i\right)\ ,
\label{Sfermi}
\end{eqnarray}
where $A_\mu$ ($\mu=0,1$) is the $U(1)$ gauge field, $\psi_i$
($i=1,\cdots, N_f$) are complex Dirac fermions of charge $k\in\Z_{>0}$.
We take the representation  $\gamma^0=\sigma^1$, $\gamma^1=i\sigma^2$
and write $\psi^i=(\psi^i_R,\psi^i_L)^T$, where $\psi_L^i$ and
$\psi_R^i$ denote the left- and right-handed components of the fermions,
respectively.
The $U(1)$ gauge transformation acts on $A=A_\mu dx^\mu$ and $\psi^i$ as
\begin{eqnarray}
A\ra A+d\lambda\ ,~~~\psi^i\ra e^{-ik\lambda} \psi^i\ ,  
\label{gauge}
\end{eqnarray}
where $\lambda$ is a $2\pi$ periodic
({\it i.e.} $\lambda$ is identified with $\lambda+2\pi$) real scalar
field.
The gauge field is normalized such that it satisfies the usual flux
quantization condition:
\begin{eqnarray}
\frac{1}{2\pi}\int F\in\Z \ ,
\label{fluxquant}
\end{eqnarray}
where $F=dA=\half F_{\mu\nu}dx^\mu dx^\nu$.
Because of this quantization condition, we are not allowed
to rescale the gauge field and the gauge coupling $e$ to eliminate the
$U(1)$ charge $k$.
In fact, as we will shortly see, vacuum structure of the 2 dim QED
with charge $k$ fermions is completely different from that with charge
$1$ fermions.

The classical global symmetry of the theory is\footnote{There are other
symmetries such as Poincar\'e symmetry, parity, time reversal and charge
conjugation, which will not be considered in this paper.}
\begin{eqnarray}
G_{\rm classical}=\frac{SU(N_f)_L\times SU(N_f)_R \times U(1)_A/\Z_2}
{(\mathbb{Z}_{N_f})_L \times (\mathbb{Z}_{N_f})_R}\ ,
\label{sym}
\end{eqnarray}
where $SU(N_f)_L$ and $SU(N_f)_R$ are the chiral symmetry that act on
$\psi_L^i$ and $\psi_R^i$, respectively, as
\begin{eqnarray}
 \psi_L^i\ra (g_{L})^i_{~j}\psi_L^j\ ,~~~ \psi_R^i\ra (g_{R})^i_{~j}\psi_R^j
\end{eqnarray}
with $(g_L,g_R)\in SU(N_f)_L\times SU(N_f)_R$,
and $U(1)_A$ is the classical axial symmetry that acts on the fermions as
\begin{eqnarray}
 \psi_L^i\ra e^{-i\alpha} \psi_L^i\ ,~~~
 \psi_R^i\ra e^{+i\alpha} \psi_R^i\ ,
\label{U1A}
\end{eqnarray}
with $e^{i\alpha}\in U(1)_A$. Note that 
$U(1)_A$ is divided by $\Z_2$ in (\ref{sym}) with the identification
$e^{i\alpha}\sim -e^{i\alpha}$, because the axial transformation
 (\ref{U1A}) with $\alpha=\pi$ corresponds
to $\psi^i\ra -\psi^i$ which is an element of the $U(1)$ gauge
transformation.
The denominator $(\mathbb{Z}_{N_f})_L \times (\mathbb{Z}_{N_f})_R$
in (\ref{sym}) acts on $(g_L,g_R,e^{i\alpha})\in
SU(N_f)_L\times SU(N_f)_R \times U(1)_A$ as
\begin{eqnarray}
(g_L,g_R,e^{i\alpha})\ra
 (\omega_Lg_L,\omega_Rg_R,e^{i\alpha}\omega_L^{1/2}\omega_R^{-1/2})
\label{ZNZN}
\end{eqnarray}
with $(\omega_L,\omega_R)\in (\mathbb{Z}_{N_f})_L \times
(\mathbb{Z}_{N_f})_R$.
Though there is a sign ambiguity in the square root
$\omega_L^{-1/2}\omega_R^{1/2}$,
it is well-defined as an element of $U(1)_A/\Z_2$.

Quantum mechanically, $U(1)_A$ is broken by the anomaly. In fact
the path integral measure of the fermions $\cD\psi\cD\ol\psi$ is
transformed by
$e^{i\alpha}\in U(1)_A$ as
\begin{eqnarray}
\cD\psi\cD\ol\psi\ra
\cD\psi\cD\ol\psi
\exp\left(-i\frac{\alpha}{\pi} k N_f\int F\right)\ ,
\label{anomaly}
\end{eqnarray}
and hence, the partition function is invariant only when
$\alpha=\pi\frac{l}{k N_f}$ with $l=1,2,\cdots 2kN_f$.
Therefore, $U(1)_A$ is
broken explicitly to $(\Z_{2k N_f})_A$ by anomaly and
the global symmetry $G$ is given by replacing
 $U(1)_A/\Z_2$ in (\ref{sym}) with
$(\Z_{2k N_f})_A/\Z_2 \equiv \Z_{kN_f}^\ax$:\footnote{
See \cite{Tanizaki:2018wtg} for similar consideration in 4 dim massless
QCD.}
\begin{eqnarray}
G=\frac{SU(N_f)_L\times SU(N_f)_R \times \mathbb{Z}_{kN_f}^\ax}
{(\mathbb{Z}_{N_f})_L \times (\mathbb{Z}_{N_f})_R}\ .
\label{symmetry}
\end{eqnarray}
Below, we will argue that $\mathbb{Z}_{kN_f}^\ax$ is spontaneously broken to
$\mathbb{Z}_{N_f}$, resulting in $k$ vacua\footnote{
The case with $N_f=1$ is studied in \cite{Anber:2018jdf}, in which case
the vacuum structure is similar to that of  4 dim ${\cal N}=1$ $SU(N)$
SYM, where $U(1)_A$ is broken to $\mathbb{Z}_{2N}$ by anomaly and
further broken spontaneously to $\mathbb{Z}_{2}$, resulting
in $N$ vacua.}.

In addition, the system admits a global $\Z_k$ 1-form symmetry,
denoted by $\Z_k^{\oneform}$.
To be explicit, we compactify the spatial direction to $S^1$
of radius $R$. Then,  the elements of $\Z_k^\oneform$ is represented
by the transformation
\begin{eqnarray}
A\ra A+\frac{1}{k}d\xi\ ,~~~
\psi^i\ra e^{-i\xi}\psi^i
\label{1formsym}
\end{eqnarray}
with a $2\pi$ periodic real scalar field $\xi$ satisfying
\begin{eqnarray}
\xi(x^0,x^1+2\pi R)=\xi(x^0,x^1)+2\pi l\ ,~~~(l=1,2,\cdots,k)
\label{xiperiod}
\end{eqnarray}
up to the gauge transformation (\ref{gauge}). When we choose $\xi=l x^1/R$,
it gives a constant shift of $A_1$ as
\begin{eqnarray}
A_1\ra A_1+\frac{l}{kR}\ .
\label{A1shift}
\end{eqnarray}
Note that this transformation should not be considered as a part of the
gauge transformation (\ref{gauge}) with $\lambda=\xi/k$, unless
$l\in k\Z$, because
(\ref{xiperiod}) is not compatible with the $2\pi$ periodicity of
$\lambda$ in (\ref{gauge}).
Under this transformation,
the Wilson loop operator that winds around the spatial circle
$W\equiv\exp\left(i\int_{S^1}A\right)$ transforms as
\begin{eqnarray}
 W\ra e^{i\frac{2\pi l}{k}}W\ .
\label{Wtr}
\end{eqnarray}

As discussed in \cite{Anber:2018jdf} for the $N_f=1$ case,
this $\Z_k^\oneform$ and the discrete axial symmetry $\Z_{kN_f}^\ax$ have
a mixed 't Hooft anomaly. Indeed, gauging $\Z_k^{\oneform}$ is equivalent
to introducing a background gauge configuration with fractional flux
quantization condition\footnote{Following \cite{Gaiotto:2017yup},
one can gauge $\sZ_k^\oneform$ as follows.
Suppose we have a theory with
a $U(1)$ gauge field $A$, which has a global $U(1)$ 1-form symmetry
given by $A\ra A+\alpha$ with $\alpha$ being a closed 1-form.
One can gauge this 1-form symmetry by promoting $\alpha$ to be
a 1-form gauge field and introducing a 2-form gauge field $B$
that transforms as $B\ra B+d\alpha$. This $U(1)$ 1-form symmetry can be
broken to $\sZ_k$ by introducing an additional 1-form gauge field $C$
that transform as $C\ra C+k\alpha$ and satisfy a constraint
$dC=kB$, which is an analogue of
a would-be Nambu-Goldstone mode of a charge $k$ Higgs field.
An action that is invariant under the $\sZ_k$ 1-form gauge symmetry
can be obtained by replacing $A$ with $\wt A\equiv A-\frac{1}{k}C$.
Then, the flux quantization condition for $\wt A$ is given by (\ref{fracflux}).
}
\begin{eqnarray}
\frac{1}{2\pi}\int F\in \frac{1}{k}\Z\ .
\label{fracflux}
\end{eqnarray}
and this makes $(\Z_{2kN_f})_A$ anomalous, because the fermion path integral
measure (\ref{anomaly}) is not invariant under generic $(\Z_{2kN_f})_A$
transformations.
The anomaly free part of $(\Z_{2kN_f})_A$ in such backgrounds
is a $\Z_{2N_f}$ subgroup, whose elements are given by
$e^{i\alpha}\in U(1)_A$ with $\alpha=\pi\frac{l}{N_f}$ with $l=1,2,\cdots,2N_f$.
Therefore, $\Z_{kN_f}^\ax(=(\Z_{2kN_f})_A/\Z_2)$ is broken to $\Z_{N_f}$
by the mixed anomaly.
Note that the unbroken subgroup $\Z_{N_f}$ of $\Z_{kN_f}^\ax$
is equivalent, under the identification (\ref{ZNZN}), to the center
$\Z_{N_f}$ of $SU(N_f)_L$ (or $SU(N_f)_R$).

\subsection{Bosonization and 't Hooft anomaly matching}

\subsubsection{$N_f=1$}
\label{oneflavor}

Let us start with the one flavour theory.
We analyze the system by using bosonization.
For a direct and rigorous argument in terms of the original
fermionic description, see \cite{Anber:2018jdf}.

It is known that the one flavour 2 dim QED can be mapped to a theory
with a $2\pi$ periodic real scalar field $\varphi$ with the
action\footnote{See e.g.\cite{Frishman:1992mr} for a review.}
\footnote{
The action (\ref{Sboson1}) is identical to the dual description of the
St\"uckelberg action studied in \cite{Komargodski:2017dmc} and we can
borrow some of the arguments given there.
}
\begin{eqnarray}
S=\int d^2x \left (-\frac{1}{4e^2} F_{\mu\nu}^2
+\frac{1}{8\pi} \partial_\mu \varphi \partial^\mu
\varphi +\frac{k}{2\pi} \varphi F_{01}
\right )\ .
\label{Sboson1}
\end{eqnarray}
The correspondence is roughly given by\footnote{
More precisely, the first relation is given as
$\psi_R^\dag\psi_L=-\frac{e^\gamma}{4\pi}\mu\cN_\mu e^{i\varphi}$,
where $\cN_\mu$ denotes the normal ordering with scale $\mu$.
See, e.g., \cite{Smilga:1992hx}.
}
\begin{eqnarray}
\psi_R^\dag\psi_L \sim c\, e^{i\varphi}\ ,~~~
\ol\psi\gamma^\mu\psi
\sim\frac{1}{2\pi}\epsilon^{\mu\nu}\del_\nu\varphi\ ,~~~
\ol\psi \gamma^3\gamma^\mu\psi
\sim\frac{1}{2\pi}\del^\mu\varphi\ ,
\label{bosonization}
\end{eqnarray}
where $c$ is a non-zero constant,
$\epsilon_{\mu\nu}$ is the anti-symmetric tensor (Levi-Civita symbol in
2 dim) with $\epsilon^{01}=-\epsilon_{01}=1$, and $\gamma^3=\sigma^3$ is
the chirality operator in 2 dim.

As discussed in section \ref{symanom}, the $U(1)_A/\Z_2$ symmetry
is broken by anomaly to $\mathbb{Z}_{k}$.
This is manifest in the bosonized language, where the classical action
(\ref{Sboson1}) captures the anomaly.
The correspondence (\ref{bosonization}) suggests that 
$e^{i\alpha}\in U(1)_A$ acts on $\varphi$ as
\begin{eqnarray}
 \varphi \ra \varphi-2\alpha\ .
\label{phishift}
\end{eqnarray}
With the generic fluxes with (\ref{fluxquant}), the action (\ref{Sboson1})
is invariant (up to $2\pi$ shifts) under this transformation only when
$2\alpha=2\pi\frac{l}{k}$ with $l=1,2,\cdots,k$, which gives $\Z_k^\ax$.

The 1-form symmetry $\Z_k^{\oneform}$ is less obvious.\footnote{
See \cite{Komargodski:2017dmc} for another explanation of the 1-form symmetry
in the bosonized theory.
}
At first sight, the action (\ref{Sboson1}) looks invariant under
any constant shift of the gauge field. In fact, one can construct
the Noether current associated with this 1-form symmetry
\begin{eqnarray}
 J_{\mu\nu}=\frac{1}{e^2}F_{\mu\nu}
-\frac{k}{2\pi}\varphi\epsilon_{\mu\nu}\ .
\label{Jmn}
\end{eqnarray}
It satisfies the conservation law
\begin{eqnarray}
 \del^\mu J_{\mu\nu}=0\ ,
\end{eqnarray}
and formally generate the phase shift of the Wilson loop operator $W$
introduced in section \ref{symanom} as
\begin{eqnarray}
e^{i\alpha J_{01}}W e^{-i\alpha J_{01}}=e^{i\alpha}W\ .
\label{Wtr2}
\end{eqnarray}
(See below for the canonical quantization of the system to show this
explicitly.)
However, this current is well-defined only modulo $k$, because $\varphi$ is
$2\pi$ periodic, and $J_{01}$ should be identified with $J_{01}+k$. This means
$e^{i\alpha J_{01}}$ is well-defined only when $\alpha=2\pi\frac{l}{k}$
($l=1,2,\cdots,k$), reproducing (\ref{Wtr}).
Thus, the generator of $\Z_k^\oneform$ is
\begin{eqnarray}
\wh U\equiv
 \exp\left(\frac{2\pi i}{k}J_{01}
\right)\ .
\label{U}
\end{eqnarray}
Note that $\wh U^k$ corresponds to a large gauge transformation
((\ref{A1shift}) with $l=k$) and the group generated by
$\wh U$ is $\Z_k$ up to gauge transformations.

Alternatively, one could start from the action
\begin{eqnarray}
S=\int d^2x \left (-\frac{1}{4e^2} F_{\mu\nu}^2
+\frac{1}{8\pi} \partial_\mu \varphi \partial^\mu
\varphi-\frac{k}{2\pi}\epsilon^{\mu\nu}\del_\mu\varphi A_\nu
\right )\ ,
\label{Sboson2}
\end{eqnarray}
which is obtained from (\ref{Sboson1}) by integrating  by parts.
In this case,  the 1-from $\Z_k^{\oneform}$ symmetry given by
$A\ra A+\frac{1}{k}d\xi$ with $d\xi$ being a closed one-form
with $2\pi$ periods is manifest,\footnote{
$\frac{1}{2\pi}\int d\varphi\wedge d\xi$ is an element of $2\pi\Z$.
} while
the axial $\Z_k^\ax$ symmetry is less obvious.
The action (\ref{Sboson2}) is invariant under any constant shift of $\varphi$
and one can construct the Noether current associated to this symmetry:
\begin{eqnarray}
J^A_\mu= \frac{1}{4\pi}\del_\mu\varphi-\frac{k}{2\pi}\epsilon_{\mu\nu} A^\nu
\end{eqnarray}
The conservation law $\del^\mu J_\mu^A=0$ follows from the equations
of motions and reproduces the anomaly equation in the fermionic theory
via the correspondence (\ref{bosonization}). However, this current
is not gauge invariant. The conserved charge
\begin{eqnarray}
 Q\equiv\int_{S^1} dx^1 J_0^{A}=
\int_{S^1} dx^1\left(\frac{1}{4\pi}\del_0\varphi-\frac{k}{2\pi} A_1
\right)
\label{Q}
\end{eqnarray}
is well-defined only modulo $k$, because a large gauge transformation
((\ref{A1shift}) with $l=k$) induces
\begin{eqnarray}
 Q\ra Q-k \ .
\end{eqnarray}
A well-defined operator can be constructed as
\begin{eqnarray}
\wh V\equiv\exp\left(-{\frac{2\pi i}{k}Q} \right)\ .
\label{V}
\end{eqnarray}
This operator gives
(\ref{phishift}) with $2\alpha=\frac{2\pi}{k}$,
and hence generates the axial $\Z_k^\ax$ symmetry.\footnote{$\wh V^k$
induces $\varphi\ra\varphi-2\pi$, which is a trivial transformation
under the identification $\varphi\sim\varphi+2\pi$.
In fact, the $2\pi$ shift of $\varphi$ can be understood as a large
gauge transformation of the 0-form gauge field $\varphi$.
We regard the transformation by $\wh V^k$ as a gauge transformation.}

It is now straightforward to check that the mixed 't Hooft anomaly of
$\Z_k^\ax$ and $\Z_k^{\oneform}$ matches with that in the fermionic
theory. If one considers a generic background with the fractional flux
quantization condition (\ref{fracflux}), the transformation
(\ref{phishift}) leaves the action (\ref{Sboson1}) invariant
 (up to $2\pi$ shifts) only when $2\alpha \in 2\pi\Z$.
This breaks $\Z_{k}^\ax$ to nothing, reproducing
the mixed 't Hooft anomaly discussed in section \ref{symanom}.

Another way of checking the mixed 't Hooft anomaly is to use
the commutation relation of the generators of 
$\Z_k^\ax$ and $\Z_k^{\oneform}$. In the $A_0=0$ gauge the
action (\ref{Sboson1}) becomes
\begin{eqnarray}
S=\int d^2x \left (\frac{1}{2e^2} (\del_0A_1)^2
+\frac{1}{8\pi} \partial_\mu \varphi \partial^\mu
\varphi +\frac{k}{2\pi} \varphi \del_0A_1
\right )\ .
\label{A0=0}
\end{eqnarray}
The canonical momenta conjugate to $A_1$ and $\varphi$ are
\begin{eqnarray}
\Pi_A\equiv \frac{1}{e^2}\del_0A_1+\frac{k}{2\pi}\varphi
=J_{01}
\ ,~~~
\Pi_\varphi\equiv \frac{1}{4\pi}\del_0\varphi\ ,
\end{eqnarray}
respectively, and the Hamiltonian is
\begin{eqnarray}
 H=\int dx^1\left(
\frac{e^2}{2}\left(
\Pi_A-\frac{k}{2\pi}\varphi
\right)^2+2\pi\Pi_\varphi^2+\frac{1}{8\pi}(\del_1\varphi)^2
\right)\ .
\label{Hami}
\end{eqnarray}
Note that the Gauss law equation (equation of motion for $A_0$) implies
\begin{eqnarray}
\del_1 \Pi_A=0\ ,
\label{GLaw}
\end{eqnarray}
which will be imposed on the physical states.

By using the canonical commutation relations,
\begin{eqnarray}
[A_1(x^0,x^1),\Pi_A(x^0,y^1)]=i\delta(x^1-y^1) \ ,~~~
[\varphi(x^0,x^1),\Pi_\varphi(x^0,y^1)]=i\delta(x^1-y^1) \ ,
\end{eqnarray}
we can explicitly check that $J_{01}$ and $Q$ introduced in (\ref{Jmn})
and (\ref{Q}) above commute with the Hamiltonian, and 
$\Z_k^{\oneform}$ and $\Z_k^\ax$ generated by $\wh V$ and $\wh U$ defined in
(\ref{V}) and (\ref{U}), respectively, are the symmetry of the system.
The important point is that $\wh U$ and $\wh V$ do not commute with each other,
but satisfy the following non-commutative relation:
\begin{eqnarray}
\wh U\wh V=\wh V\wh U e^{\frac{2\pi i}{k}}\ .
\label{NC}
\end{eqnarray}
This relation follows from the commutation relation\footnote{
See \cite{Anber:2018jdf} for the derivation without using bosonization.
See also \cite{Anber:2018xek} for the realization of this algebra in
a TQFT describing the IR physics of the system.}
\begin{eqnarray}
[J_{01}, Q]
=\left[\Pi_A,\int dx^1\left(\Pi_\varphi-\frac{k}{2\pi}A_1\right)\right]
=i\frac{k}{2\pi}\ .
\end{eqnarray}
Therefore, if one promotes the transformation by $\wh U$ to
a gauge symmetry, the operator $\wh V$ is no longer gauge invariant,
which means that when $\Z_k^\oneform$ is gauged,
$\Z_k^\ax$ is not a symmetry of the system any more.
This is consistent with the mixed 't Hooft anomaly discussed above.

\subsubsection{$N_f>1$}
\label{multi}

Let us next discuss the generalization to the multi flavour theory.
The multi flavour theory can be bosonized by using the non-Abelian 
bosonization rules \cite{Witten:1983ar} (See e.g. \cite{Frishman:1992mr}
for a review.)
\begin{eqnarray}
\psi_{Rj}^\dag\psi_L^i\sim c\,u^i_{~j}\ ,~~~
J_{-j}^i\sim\frac{i}{2\pi}(u\del_- u^{-1})^i_{~j}\ ,~~~
J_{+j}^i
\sim\frac{i}{2\pi}(u^{-1}\del_+ u)^i_{~j}\ ,
\label{bosonization2}
\end{eqnarray}
where $u$ is a $U(N_f)$ valued
scalar field, $(J_-,J_+)$ are the $U(N_f)_L\times U(N_f)_R$ currents,
and $\del_\pm\equiv \frac{1}{\sqrt{2}}(\del_0\pm\del_1)$.

We parametrize $u$ as
\begin{eqnarray}
 u=e^{i\varphi} g \ ,
\end{eqnarray}
where $\varphi$ is a $2\pi$ periodic real scalar field
and $g$ is an $SU(N_f)$ valued scalar field. Since $u$ is invariant
under the following $\Z_{N_f}$ transformation
\begin{eqnarray}
 \varphi\ra\varphi-\frac{2\pi}{N_f}\ ,~~~g\ra e^{\frac{2\pi i}{N_f}}g\ ,
\label{ZNfiden}
\end{eqnarray}
the fields $(\varphi,g)\in U(1)\times SU(N_f)$ related by this
transformation are identified. We regard this
$\Z_{N_f}$ symmetry as a gauge symmetry.

The bosonization rules (\ref{bosonization2}) imply that
the $U(1)_V$ current $J_\mu^V=\tr J_\mu$ is given by
\begin{eqnarray}
 J_\mu^V\sim\frac{N_f}{2\pi}\epsilon_{\mu\nu}\del^\nu\varphi\ ,
\end{eqnarray}
and hence only $\varphi$ couples with the gauge field.
The action is given by
\begin{eqnarray}
S=\int d^2x \left (-\frac{1}{4e^2} F_{\mu\nu}^2
+\frac{N_f}{8\pi} \partial_\mu \varphi \partial^\mu
\varphi +\frac{kN_f}{2\pi} \varphi F_{01}
\right )
+S_{\rm WZW}(g)\ ,
\label{Sboson-multi}
\end{eqnarray}
where $S_{\rm WZW}(g)$ is the action of the $SU(N_f)$ WZW theory at level 1.

It is not difficult to check that the global symmetry of the system agrees with
that of 2 dim QED discussed in section \ref{symanom}. From the correspondence
(\ref{bosonization2}), we find that $(g_L,g_R)\in SU(N_f)_L\times SU(N_f)_R$
acts on $g$ as
\begin{eqnarray}
 g\ra g_L g g_R^{-1}\ ,
\label{gLggR}
\end{eqnarray}
and the generator of $\Z_{kN_f}^\ax$ acts on $\varphi$ as
\begin{eqnarray}
 \varphi\ra\varphi-\frac{2\pi}{kN_f}\ .
\label{axialshift}
\end{eqnarray}
One can also check that
the denominator $(\Z_{N_f})_L\times(\Z_{N_f})_R$ in (\ref{symmetry})
acts trivially on $\varphi$ and $g$ under the identification
by (\ref{ZNfiden}).
Repeating the same argument as in section \ref{oneflavor} for the $N_f=1$
case with the identification  (\ref{ZNfiden}),
we find that the one form symmetry is $\Z_k^{\oneform}$ as expected.

The 't Hooft anomaly matching for the mixed anomaly of
$\Z_{k N_f}^\ax$ and $\Z_k^\oneform$ works as well. When we allow
the fractional flux (\ref{fracflux}), the unbroken part of the
$\Z_{kN_f}^\ax$ shift symmetry (\ref{axialshift}) becomes $\Z_{N_f}$
generated by $\varphi\ra\varphi-\frac{2\pi}{N_f}$. This precisely agrees
with the mixed 't Hooft anomaly discussed in section \ref{symanom}.

The discussion below (\ref{A0=0}) can also be applied to the $N_f>1$
cases. Generalizing (\ref{U}) and (\ref{V}), the operators that generate
$\Z_k^\oneform$ and $\Z_{kN_f}^\ax$ are obtained as
\begin{eqnarray}
\wh U&\equiv&
 \exp\left(\frac{2\pi i}{k}J_{01}\right)
=
 \exp\left(\frac{2\pi i}{k}\Pi_A\right)\ ,
\label{U2}
\\
\wh V&\equiv&
 \exp\left(-{\frac{2\pi i}{k}Q} \right)
=
 \exp\left(-\frac{2\pi i}{kN_f}
\int dx^1\Pi_\varphi
+i\int dx^1 A_1
\right)\ ,
\label{V2}
\end{eqnarray}
respectively, where
\begin{eqnarray}
\Pi_A\equiv \frac{1}{2e^2}\del_0A_1
+\frac{kN_f}{2\pi}\varphi\ ,~~~
\Pi_\varphi\equiv\frac{N_f}{4\pi}\del_0\varphi
\end{eqnarray}
are the canonical momenta conjugate to $A_1$ and $\varphi$,
respectively. They also satisfy (\ref{NC}).
As discussed in section \ref{oneflavor},
when the transformation by $\wh U$ is gauged, $\wh V$ is no longer
a gauge invariant operator. But, in this case, $\wh V^k$ commutes with $\wh U$
and hence it is well-defined. Since $\wh V^k$ generates the $\Z_{N_f}$
subgroup of $\Z_{kN_f}^\ax$, we conclude that $\Z_{kN_f}^\ax$ is broken
to $\Z_{N_f}$ when $\Z_k^\oneform$ is gauged.
This is again consistent with the mixed 't Hooft anomaly discussed above.

\subsection{Spontaneous symmetry breaking and the vacuum structure}
\label{SSB}

The existence of the 't Hooft anomaly already predicts that the vacuum
has to be non-trivial. In fact, because the space of ground states
has to be a representation of the algebra generated by $\wh U$ and
$\wh V$ satisfying (\ref{NC}), the vacua have to be at least $k$-fold
degenerate.
In this subsection, we construct the ground states explicitly
and argue that $\Z_{kN_f}^\ax$
is spontaneously broken down to $\Z_{N_f}$ via the condensate
\begin{eqnarray}
 \VEV{\det{(\psi_{Rj}^\dag\psi^i_L)}}\ne 0\ ,
\label{VEVdet}
\end{eqnarray}
which suggests that there are indeed $k$ degenerate vacua
(apart from the $\theta$ vacua)
in 2 dim massless QED with $N_f$ flavours.

\subsubsection{$N_f=1$}
\label{SSBNf1}

In order to see the vacuum structure more explicitly,
it is again useful to start with the $N_f=1$ case.
For the one flavour 2 dim QED, as it is well-known in the $k=1$ case
\cite{Coleman:1976uz}
and recently shown for general $k$ in \cite{Anber:2018jdf}, the fermion bilinear
operator $\psi_R^\dag\psi_L$ has a non-zero vacuum expectation value. As a result,
$\Z_{k}^\ax$ is spontaneously broken to nothing and there are $k$ vacua
associated with this breaking.
The axial symmetry $\Z_{k}^\ax$ acts as a cyclic rotation of
these $k$ vacua. From the anomaly relation (\ref{anomaly}), this fact implies
that the $i$-th vacuum is mapped to the $(i+1)$-th vacuum (mod $k$) when the
$\theta$ angle is shifted by $2\pi$.

Let us explain how the spontaneous breakdown of $\Z_k^\ax$ can be
understood in terms of the bosonized theory (\ref{Sboson1}).\footnote{
See \cite{Komargodski:2017dmc} for another closely related derivation
of the $k$-fold degeneracy of the ground state in the bosonized
theory (\ref{Sboson1}).}
To this end, let us parametrize the ground states $\ket{\theta}$
by the eigenvalue $\theta\in\R$ of the operator
$2\pi \Pi_A$.\footnote{
Here we assume that $\ket{\theta}$ is the unique vacuum with
$2\pi\Pi_A\ket{\theta}=\theta\ket{\theta}$ and see that this assumption
is consistent with (\ref{NC}).
}
This is possible because $\Pi_A$ commutes with the
Hamiltonian. Note also that $\theta$ is a constant
because of the Gauss law equation (\ref{GLaw}).
The operator $\wh U$
defined in (\ref{U2}) acts on $\ket{\theta}$ as\footnote{
$\theta$ used in \cite{Anber:2018jdf} is $\theta/k$ in our notation.
}
\begin{eqnarray}
\wh U\ket{\theta}= e^{i\frac{\theta}{k}}\ket{\theta}\ .
\label{Uvac}
\end{eqnarray}
The normalization of $\theta$ has been chosen such that the eigenvalue
of the operator $\wh U^k$, which corresponds to the large gauge
transformation, is $e^{i\theta}$.
\footnote{
Although we have regarded $\wh U$ as the generator of $\Z_k^\oneform$,
it doesn't satisfy $\wh U^k=1$ on $\ket{\theta}$. It is easy to fix this
by defining an operator $\wh U'\equiv e^{-i\frac{\theta}{k}}\wh U$
that satisfies $\wh U'^k=1$ on the subspace of the Hilbert space we are
interested in without affecting the non-commutative relation (\ref{NC}).
See (\ref{clockshift}).
}
This is compatible with
the phase factor induced by the $\theta$ term
$\frac{\theta}{2\pi}\int F$ and this parameter $\theta$ is identified
as the $\theta$ angle in 2 dim QED.
Because $\wh V$ (defined in (\ref{V2}) with $N_f=1$) satisfies
the relation $\wh V\Pi_A\wh V^{-1}=\Pi_A-1$, we find
\begin{eqnarray}
 \wh V\ket{\theta}=\ket{\theta+2\pi}\ .
\label{Vvac}
\end{eqnarray}
It is easy to see that (\ref{Uvac}) and (\ref{Vvac}) are compatible
with the relation (\ref{NC}).

On the ground state $\ket{\theta}$, the Hamiltonian (\ref{Hami}) looks like that of a free massive scalar field with a potential
\begin{eqnarray}
 V(\varphi)=\frac{e^2}{8\pi^2}\left(k\varphi-\theta\right)^2\ .
\end{eqnarray}
However, one should be aware that $\varphi$ is a $2\pi$ periodic
scalar field. In other words, $\wh V^k$ is regarded as a gauge symmetry.
Therefore, we should mod out the system by the transformation
\begin{eqnarray}
\varphi\ra\varphi-2\pi n\ ,~~~
\ket{\theta}\ra\ket{\theta+2\pi kn}\ ,~~~(n\in\Z)\ .
\label{varphishift}
\end{eqnarray}
One way to achieve this is to pick $\ket{\theta}$
as a representative of the equivalence class
$\big\{\ket{\theta+2\pi kn}|\,n\in\Z\big\}$
and regard $\varphi$ as a non-compact scalar field that
takes values in $\R$ without any identifications.
With this understanding,
we can regard the system as the theory of
a free massive real scalar field with the Hamiltonian given by
setting $\Pi_A=\frac{\theta}{2\pi}$ in (\ref{Hami}):
\begin{eqnarray}
 H=\int dx^1\left(
2\pi\Pi_\varphi^2+\frac{1}{8\pi}(\del_1\varphi)^2
+\frac{e^2}{8\pi^2}(k\varphi-\theta)^2
\right)\ ,
\label{Hami2}
\end{eqnarray}
when we only consider the vacuum $\ket{\theta}$ and gauge invariant
local operators acting on it, {\it i.e.} the superselection sector
constructed on $\ket{\theta}$.

In fact, in various literature on 2 dim QED (with $N_f=1$ and $k=1$),
this Hamiltonian (\ref{Hami2}) is used as the starting point of the
bosonized description.
One new feature that appears in the $k>1$ case is that there are
$k$ sectors, $\big\{\ket{\theta+2\pi(kn+j)}|\,n\in\Z\big\}$ labeled by
$j=0,1,\cdots,k-1$ ($\mod k$), that are not gauge equivalent.
These $k$ sectors are related to each other by the action of $\Z_k^\ax$
generated by $\wh V$. This means that the $\Z_k^\ax$ is spontaneously broken
down and there are $k$ vacua related by $\Z_k^\ax$, or, equivalently,
 $2\pi$ shifts of the parameter $\theta$.
The theory is invariant under the $2\pi$ shift of
$\theta$, but in order to come back to the original vacuum,
$\theta$ has to be shifted by $2\pi k$.
These properties cannot be seen if one starts with the system
defined by (\ref{Hami2}).

Note also that the Hamiltonian (\ref{Hami2})
cannot be used if one wants to consider operators that do not
commute with $\wh U$, because such operators
change the eigenvalue of $\wh U$ and map a state to
a different sector.  The Wilson loop operator $W$
that winds once around the spatial circle is such an example.
It changes the eigenvalue of $\wh U$ by a factor of $e^{\frac{2\pi i}{k}}$.
In general, there is no gauge invariant operator that
changes the eigenvalue of $\wh U$ by a factor
other than $e^{\frac{2\pi i}{k}n}$ with $n\in\Z$, because
the gauge invariant operator should commute with the generator of the
large gauge transformation $\wh U^k$.
Therefore, we may restrict
the whole Hilbert space to the subspace that is constructed
by acting gauge invariant operators on the $k$ vacua represented by
$\ket{\theta+2\pi j}$ ($j=0,1,\cdots, k-1$) with fixed $\theta$.
The operators $\wh U$ and $\wh V$ are represented on the
$k$ dimensional vacuum space by the clock and shift matrices as
\begin{eqnarray}
\wh U'\equiv e^{-i\frac{\theta}{k}} \wh U=
\left(
\begin{array}{cccc}
1\\
&\omega\\
&&\ddots\\
&&&\omega^{k-1}
\end{array}
\right)
\ ,~~~
\wh V=
\left(
\begin{array}{cccc}
&&&1\\
1\\
&\ddots\\
&&1
\end{array}
\right)\ ,
\label{clockshift}
\end{eqnarray}
where $\omega\equiv e^{\frac{2\pi i}{k}}$.

We can also argue the spontaneous breakdown of $\Z_k^\ax$
by using order parameters characterizing it.
(\ref{NC}) implies that the operator $\wh U$ transforms as
\begin{eqnarray}
 \wh U\ra \wh V\wh U \wh V^{-1}=e^{-\frac{2\pi i}{k}}\wh U
\end{eqnarray}
by the generator of $\Z_k^\ax$. Therefore, the fact that
$\bra{\theta}\wh U\ket{\theta}=e^{i\frac{\theta}{k}}$ is non-zero
implies the breaking of $\Z_k^\ax$.
Another natural order parameter is the vacuum expectation value of
$e^{i\varphi}$, which is related to $\psi_R^\dag\psi_L$ by the
bosonization rule (\ref{bosonization}).
Since the minimum of the potential in (\ref{Hami2})
is $\varphi=\theta/k$, the classical value is
\begin{eqnarray}
\bra{\theta}e^{i\varphi}\ket{\theta}= e^{i\frac{\theta}{k}} \ .
\label{VEVeiphi}
\end{eqnarray}
Quantum mechanically, the operator $e^{i\varphi}$ should be defined
by taking the normal ordering. Because $\varphi$ in (\ref{Hami2}) is a massive
free scalar field, this can be easily done and the result (\ref{VEVeiphi})
is unchanged, which again shows the breaking of $\Z_k^\ax$.

\subsubsection{$N_f>1$}
\label{SSBNf>1}

Since the non-Abelian part involving the $SU(N_f)$ valued field
 $g$ in (\ref{Sboson-multi}) decouples from $\varphi$ and the
gauge field, most of the arguments in section \ref{SSBNf1} go through with a
little modification.
The main difference is that $(\varphi,g)$ is identified by
the transformation (\ref{ZNfiden}) rather than a simple
$2\pi$ shift of $\varphi$. The transformation (\ref{ZNfiden}) is induced by
the operator $\wh V^k \wh\omega_L$, where $\wh\omega_L$ is the operator
that induces the transformation  (\ref{gLggR}) with
$g_L=e^{\frac{2\pi i}{N_f}}$.
Therefore, this operator $\wh V^k \wh\omega_L$ is the generator
of the $\Z_{N_f}$ gauge symmetry (\ref{ZNfiden})
and plays the same role as $\wh V^k$
in the $N_f=1$ case in section \ref{SSBNf1}.

A crucial point is that, because of the Coleman-Mermin-Wagner
theorem,\cite{Coleman:1973ci,Mermin:1966fe,Hohenberg:1967zz} the continuous
chiral symmetry $SU(N_f)_L\times SU(N_f)_R$ cannot be spontaneously broken.
This means that the ground states are the singlet state as a representation
of $SU(N_f)_L\times SU(N_f)_R$ and vacuum expectation values of any
operators that are not invariant under $SU(N_f)_L\times SU(N_f)_R$ have
to vanish.
Therefore, we can basically forget about the field $g$ for the
consideration of the vacuum structure. Then, as in the $N_f=1$ case
discussed in section \ref{SSBNf1}, the vacuum is parametrized by
$\theta$ satisfying (\ref{Uvac}) and (\ref{Vvac}). This also implies
that the action of the operator $\wh V^k\wh\omega_L$
on the ground states is equivalent to the action of $\wh V^k$,
which is the generator of the $\Z_{N_f}$ subgroup of the discrete axial
symmetry $\Z_{kN_f}^\ax$. Then, we should identify the system by the
transformation
\begin{eqnarray}
\varphi\ra\varphi-\frac{2\pi n}{N_f}\ ,~~~
\ket{\theta}\ra\ket{\theta+2\pi kn}\ ,~~~(n\in\Z)\ .
\label{varphishift2}
\end{eqnarray}
that generalizes (\ref{varphishift}) to the cases with $N_f>1$.
Therefore, under this identification, $\Z_{kN_f}^\ax$ is spontaneously
broken to $\Z_{N_f}$, because the vacuum $\ket{\theta}$ is not invariant
under the action of $\wh V$ (the generator of $\Z_{kN_f}^\ax$),
but it is invariant under $\wh V^k$  (the generator of $\Z_{N_f}
\subset \Z_{kN_f}^\ax$).
Just as in the $N_f=1$ case, there are $k$ vacua represented by
$\ket{\theta+2\pi j}$ with $j=0,1,2,\cdots,k-1$ ($\mod k$),
which are related by the $2\pi$ shift of the parameter $\theta$.
Then, generalizing (\ref{Hami2}), Hamiltonian for the superselection
sector constructed on $\ket{\theta}$ with $\varphi$ regarded as a
non-compact scalar field is obtained as
\begin{eqnarray}
 H=\int dx^1\left(
\frac{2\pi}{N_f}\Pi_\varphi^2+\frac{N_f}{8\pi}(\del_1\varphi)^2
+\frac{e^2}{8\pi^2}(kN_f\varphi-\theta)^2
\right)+H_{\rm WZW}\ ,
\label{Hboson}
\end{eqnarray}
where $H_{\rm WZW}$ is the Hamiltonian for the $SU(N_f)$
valued field $g$
induced from $S_{\rm WZW}(g)$ in (\ref{Sboson-multi}).

As mentioned above, the Coleman-Mermin-Wagner theorem
implies that $SU(N_f)_L\times SU(N_f)_R$ non-singlet
operators have vanishing vacuum expectation values and hence an order
parameter that characterizes the vacuum structure
has to be the vacuum expectation value of
an $SU(N_f)_L\times SU(N_f)_R$ singlet operator.
In addition, such an operator has to be invariant under (\ref{ZNfiden}).
As discussed in section \ref{SSBNf1}, the vacuum expectation value
of the operator $\wh U$ is one of the order parameters
that shows the breaking of $\Z_{kN_f}^\ax$ to $\Z_{N_f}$.
Another natural operator is $e^{iN_f\varphi}$.
The argument around (\ref{VEVeiphi}) implies
\begin{eqnarray}
\bra{\theta}e^{iN_f\varphi}\ket{\theta}= e^{i\frac{\theta}{k}} \ ,
\label{VEVNfephi}
\end{eqnarray}
which again shows the breaking of $\Z_{kN_f}^\ax$ to $\Z_{N_f}$.
Since $e^{iN_f\varphi}=\det u$, this operator is related by
the bosonization rules (\ref{bosonization2}) to $\det(\psi_{Rj}^\dag\psi_L^i)$.
Therefore, (\ref{VEVNfephi}) suggests (\ref{VEVdet}), even though
the vacuum expectation value of
$\psi_{Rj}^\dag\psi_L^i$ vanishes for $N_f>1$.

\subsection{Mass deformation}
\label{MassDeform}

In this subsection, we consider adding a fermion mass term 
$M_0\ol\psi_j\psi^j$ to the action (\ref{Sfermi}) and
study the chiral condensate, vacuum energy and string tension for the
cases with general $N_f$ and $k$. 
 As we will see, the effect of non-zero $M_0$
drastically alters the qualitative features of the vacuum structure
for $N_f>1$ considered in section \ref{SSB}.
In the bosonized description, an interaction term proportional to
$M_0\tr(u+u^\dag)$, which corresponds to the fermion mass term via
the bosonization rule (\ref{bosonization2}), is added and the system
is no longer exactly solvable. However, we are able to get quite
non-trivial results thanks to various powerful techniques developed
in 2 dim QFT. The results in this subsection will be used in section
\ref{AppString} to give non-perturbative predictions in string theory.
In this subsection, we focus on the small mass (strong coupling) regime
$M_0\ll e$ and the large volume limit $M_0R\ra\infty$.
We also assume $M_0>0$ throughout this paper.

\subsubsection{Chiral condensate and vacuum energy}

As mentioned in section \ref{SSBNf>1}, the chiral condensate
$\VEV{\ol\psi_i\psi^j}$ vanishes for $N_f>1$
due to the Coleman-Mermin-Wagner theorem in the massless limit.
When the mass term $M_0\ol\psi_j\psi^j$ is added,
however, the continuous chiral symmetry
$SU(N_f)_L\times SU(N_f)_R$ is explicitly broken to the
diagonal $SU(N_f)$ subgroup and there is no reason for the chiral
condensate to vanish. In fact, it was shown in
\cite{Smilga:1992hx,Hetrick:1995wq,Hetrick:1995yx} that
the trace part of the chiral condensate is non-zero and behaves as
\begin{eqnarray}
 \VEV{\ol\psi_j\psi^j}\propto e^{\frac{2}{N_f+1}}M_0^{\frac{N_f-1}{N_f+1}}\ .
\label{psibarpsi}
\end{eqnarray}
This is clearly a non-perturbative effect,
since it has fractional powers of $e$ and $M_0$.

Let us first outline the derivation of (\ref{psibarpsi}) using the bosonized
Hamiltonian (\ref{Hboson}).
It is convenient to redefine the scalar field $\varphi$
as
\begin{eqnarray}
h\equiv \half\sqrt{\frac{N_f}{\pi}}\left(\varphi-\frac{\theta}{kN_f}
\right) \ ,
\end{eqnarray}
 so that  the kinetic term is canonically normalized and
the mass term for $h$ is simplified.
The Hamiltonian is given by
\begin{eqnarray}
 H=\int dx^1
\cN_\mu\left[
\half\Pi_{h}^2+\half(\del_1 h)^2
+\frac{m_h^2}{2}h^2
\right]+H_{\rm WZW}+H_{\rm mass}\ ,
\label{Hboson2}
\end{eqnarray}
where
\begin{eqnarray}
m_h^2\equiv\frac{e^2k^2N_f}{\pi}\ ,
\end{eqnarray}
and
\begin{eqnarray}
 H_{\rm mass}\equiv
\int dx^1 \mu \cN_\mu\left[
\tilde c M_0\,e^{i\left(2\sqrt{\frac{\pi}{N_f}}h+\frac{\theta}{kN_f}\right)}
\tr(g)+{\rm h.c.}
\right]\ .
\label{Hmass}
\end{eqnarray}
Here, we have regularized the system by taking the normal ordering
at scale $\mu$\footnote{See \cite{Coleman:1974bu} for the normal-ordering
prescription.}, denoted by the symbol $\cN_\mu$.
The last term $H_{\rm mass}$ in (\ref{Hboson2}) comes from
the fermion mass term via the bosonization relation\footnote{
See, e.g., \cite{Frishman:1992mr} for a review.}
\begin{eqnarray}
\psi_{Rj}^\dag\psi_L^i=\tilde c\,\mu\cN_\mu u^i_{~j}
=\tilde c\,\mu\cN_\mu e^{i\varphi}g^i_{~j}\ ,
\end{eqnarray}
where $\tilde c$ is a numerical constant.
In this section, we are interested in the small mass (strong coupling) regime
$M_0\ll e$ and treat $H_{\rm mass}$ as a small deformation of the Hamiltonian
for the $M_0=0$ case used in the previous subsections.
In this case, as we can see from (\ref{Hboson2}) and (\ref{Hmass}),
the scalar field $h$ is much heavier than $g$, and we first try to integrate
it out.
Using the formula \cite{Coleman:1974bu}:
\begin{eqnarray}
\cN_m[e^{i\beta\phi}]=\left(\frac{\mu^2}{m^2}\right)^{\frac{\beta^2}{8\pi}}
\cN_\mu[e^{i\beta\phi}]\ ,
\end{eqnarray}
where $\phi$ denotes a canonically normalized free scalar field,
and the fact that if the mass of $\phi$ is $m$,
vacuum expectation value of the left hand side is $1$, we obtain
\begin{eqnarray}
\bra{\theta}\cN_\mu\left[e^{i\,2\sqrt{\frac{\pi}{N_f}}h}\right]\ket{\theta}
=\left(\frac{m_h}{\mu}\right)^{\frac{1}{N_f}}\ . 
\label{VEVexph}
\end{eqnarray}
Replacing the operator $\cN_\mu\left[e^{i\,2\sqrt{\frac{\pi}{N_f}}h}\right]$
with its vacuum expectation value (\ref{VEVexph}), we obtain the low energy
effective Hamiltonian for the light field $g$:
\begin{eqnarray}
 H^{\rm low}_{\rm mass}=\int dx^1 \mu^{1-\frac{1}{N_f}}m_h^{\frac{1}{N_f}}
 \cN_\mu\left[ \tilde cM_0
\,e^{i\frac{\theta}{kN_f}}
\tr(g)+{\rm h.c.}
\right]\ .
\end{eqnarray}
We find that the mass scale of the light field $g$ is given by
\begin{eqnarray}
m_l^2=\mu^{1-\frac{1}{N_f}}m_h^{\frac{1}{N_f}} M_0\ .
\end{eqnarray}
Choosing the normal ordering scale as $\mu=m_l$, we get
\begin{eqnarray}
 m_l^2=m_h^{\frac{2}{N_f+1}}M_0^{\frac{2N_f}{N_f+1}}\ ,
\end{eqnarray}
and
\begin{eqnarray}
 H^{\rm low}_{\rm mass}=\int dx^1 m_l^2
 \cN_{m_l}\left[ \tilde c\,e^{i\frac{\theta}{kN_f}}
\tr(g)+{\rm h.c.}
\right]\ .
\label{Hgmass}
\end{eqnarray}
{}From this expression, we find that $m_l$ is the only mass scale of the low
energy effective Hamiltonian in the $R\ra\infty$ limit.
Then, a simple dimensional analysis implies
\begin{eqnarray}
 M_0\bra{\theta}\ol\psi_j\psi^j\ket{\theta}
=f(\theta)\, m_l^2
=f(\theta)\,m_h^{\frac{2}{N_f+1}}M_0^{\frac{2N_f}{N_f+1}}
\label{psibarpsi-ftheta}
\end{eqnarray}
with some function of $\theta$ denoted by $f(\theta)$, reproducing
(\ref{psibarpsi}).

The function $f(\theta)$ in (\ref{psibarpsi-ftheta})
was calculated in \cite{Hetrick:1995wq,Hetrick:1995yx} for the $k=1$
case. (See Appendix \ref{AppCC}.)
Since $k$ always appears in the combination $\theta/k$ or $m_h$ in the
Hamiltonian (\ref{Hboson2}), the $k$ dependence of the function
$f(\theta)$ can be included simply by rescaling $\theta\ra\theta/k$ in
the expression for $k=1$ and we obtain
\begin{eqnarray}
f(\theta)=-\frac{N_f}{4\pi}
\left(2\exp(\gamma)\cos
\left(
\frac{1}{N_f}\ol{(\theta/k)}
\right)\right)^{\frac{2N_f}{N_f+1}}
\ ,
\label{ftheta}
\end{eqnarray}
where $\gamma\simeq 0.577$ is the Euler's constant and
\begin{eqnarray}
 \ol{x}\equiv x-2\pi\left[\frac{x+\pi}{2\pi}\right]
\label{olx}
\end{eqnarray}
with $[x]$ being the floor function
that gives the greatest integer less than or equal to $x$.
Note that (\ref{olx}) implies $\ol{x}=x$ for $-\pi\le x<\pi$
and $\ol{x+2\pi}=\ol{x}$. Therefore, the expression
(\ref{ftheta}) is manifestly invariant under the $2\pi k$ shift of
$\theta$, though it has cusp singularities at $\theta/k=\pi$
($\mod 2\pi$) for $N_f>1$.\footnote{
These cusp singularities exist even for $k=1$. The existence of
the cusps can be understood from the mixed anomaly between
the vector-like flavour symmetry $SU(N_f)_V/(\Z_{N_f})_V$
and the charge conjugation symmetry discussed in
\cite{Komargodski:2017dmc}. We thank the anonymous
referee for pointing this out to us.} The $2\pi k$ periodicity 
of $\theta$ follows from the fact that $2\pi k$ shift of $\theta$ can
be absorbed by the redefinition of $g$ as
$g\ra e^{-\frac{2\pi i}{N_f}}g$ in (\ref{Hgmass}).

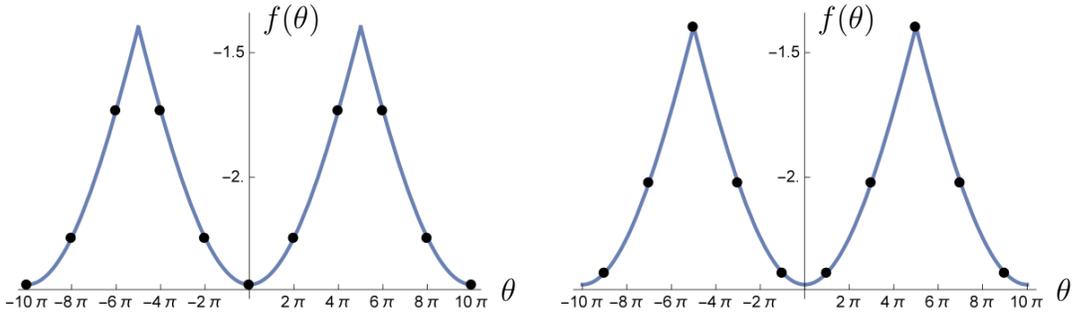
\begin{figure}[ht]
\begin{center}
\begin{picture}(420,130)(0,0)
\put(0,0){\includegraphics[scale=0.7,bb=0 0 260 161]{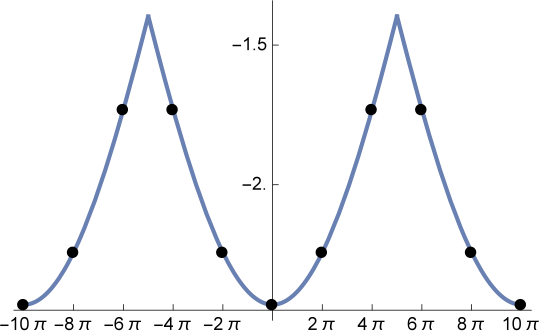}}
\put(210,0){\includegraphics[scale=0.7,bb=0 0 260 161]{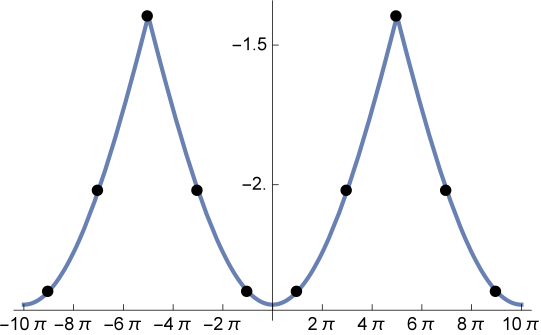}}
\put(190,8){\makebox(0,0){$\theta$}}
\put(108,110){\makebox(0,0){$f(\theta)$}}
\put(400,8){\makebox(0,0){$\theta$}}
\put(318,110){\makebox(0,0){$f(\theta)$}}
\end{picture}
\parbox{80ex}{
\caption{A plot of the function $f(\theta)$ for $k=5$ and $N_f=4$
in the range $-2\pi k\le\theta\le 2\pi k$.
 The dots in the left and right panels
are the points at $\theta=2\pi n$ and $\theta=2\pi n+\pi$
($n\in\Z$), respectively. These dots are proportional to the
expectation value of the energy density $\cE(\theta+2\pi n)$
for the states $\ket{\theta+2\pi n}$ ($n\in\Z$) with $\theta=0$ (left)
and $\theta=\pi$ (right). The $Q$-string tension $\sigma(Q)$ 
(with $Q=n\in\Z$) is proportional to the height of the dots
measured from the lowest one.
}
\label{fig1}
}
\end{center}
\end{figure}

So far, we have implicitly assumed that $\ket{\theta}$ is the vacuum state
of the system. However, this is not always true.
As we have seen in section \ref{SSB}, there are $k$ degenerate ground states
$\ket{\theta+2\pi j}$ with $j=0,1,2,\cdots,k-1$ ($\mod k$) for the $M_0=0$ case.
When the fermion mass $M_0$ is turned on, the degeneracy is lifted, because
the discrete axial symmetry $\Z_{kN_f}^\ax$ is explicitly broken by
$H_{\rm mass}$.
In fact, it can be shown that the energy density (expectation value of the
Hamiltonian density $\cH$) is proportional to the chiral condensate
(\ref{psibarpsi-ftheta}) as
\begin{eqnarray}
\cE(\theta)\equiv
\bra{\theta}\cH\ket{\theta}=
\frac{N_f+1}{2N_f}f(\theta)\, m_h^{\frac{2}{N_f+1}}M_0^{\frac{2N_f}{N_f+1}}
\ ,
\label{energydensity}
\end{eqnarray}
up to some irrelevant terms that are independent of
$\theta$ and $M_0$.\footnote{
The overall factor in (\ref{energydensity}) is different from the
expression for the energy density given in \cite{Rodriguez:1996zj}.
In \cite{Rodriguez:1996zj}, only the
contribution from the fermion mass term is taken into account. We found
that the kinetic term also has a contribution of the same order and
included in (\ref{energydensity}). See Appendix \ref{AppCC} for details.
}
(See Appendix \ref{AppCC}.)
Note that comparing (\ref{psibarpsi-ftheta}) and (\ref{energydensity}),
we get a relation
\begin{eqnarray}
\frac{\del \cE(\theta)}{\del M_0}=\bra{\theta}\ol\psi_j\psi^j\ket{\theta}\ .
\label{EM0rel}
\end{eqnarray}
The expression for the energy density (\ref{energydensity}) implies that
$\ket{\ol\theta}$ (up to the identification
 $\ket{\ol\theta}\sim \ket{\ol\theta+2\pi k j}$ ($j\in\Z$))
is the lowest energy state among the $k$ states $\ket{\theta+2\pi j}$
with $j=0,1,\cdots,k-1$ for generic $\theta$.
For $\theta=\pi$ ($\mod 2\pi$), the lowest energy states
are two-fold degenerate and given by $\ket{\pm \pi}$.
(See Figure \ref{fig1}.)
Then, the vacuum energy density
$\cE_{\rm vac}(\theta)$ is given by
\begin{eqnarray}
 \cE_{\rm vac}(\theta)=\min_{j\in\sZ}\cE(\theta+2\pi j)=\cE(\ol\theta)\ ,
\end{eqnarray}
which is a $2\pi$ periodic function with cusps at $\theta=\pi$
($\mod 2\pi$).\footnote{These cusps can be understood from the mixed
anomaly between the 1-form symmetry and the charge conjugation
symmetry.}

\subsubsection{String tension}
\label{StrTen}

The formula (\ref{energydensity}) can be used to obtain the string
tension as it was done in \cite{Rodriguez:1996zj} for $k=1$.
Let us consider an electric flux created by a pair of external point
particles of charge $\pm Q$ placed at $x^1=\mp \infty$.
We call it a $Q$-string, though it fills up the 2 dim space-time.
This amounts to adding
\begin{eqnarray}
S_{\rm int}
= Q \int dx^0 A_0|_{x^1=-\infty}-  Q\int dx^0 A_0|_{x^1=+\infty}
= Q \int F
\label{Sint}
\end{eqnarray}
in the action, which is equivalent to shifting $\theta$ to $\theta+2\pi Q$.
Therefore, the tension $\sigma(Q)$ of the $Q$-string is estimated as
\begin{eqnarray}
\sigma(Q)= \cE(\theta+2\pi Q)-\cE(\theta)\ ,
\label{tension}
\end{eqnarray}
where we have assumed $-\pi<\theta<\pi$ so that $\ket{\theta}$ is
the vacuum state.
Though (\ref{tension}) can be formally used for $Q\notin\Z$,
since the last expression in (\ref{Sint}) is manifestly gauge invariant
for any $Q\in\R$,
we impose $Q\in\Z$ so that the external point particles
are also consistent with the $U(1)$ gauge symmetry.
For $k=1$, this restriction is too strong and one always gets
$\sigma(Q)=0$, because of the $2\pi$ periodicity of the function
$\cE(\theta)$. For $k>1$, however, the periodicity of $\cE(\theta)$
becomes $2\pi k$ and we get non-trivial results for $Q=1,2,\cdots,k-1$
($\mod k$). (See Figure \ref{fig1}.)
The $Q$-strings with $Q=nk$ ($n\in\Z$) are tensionless, {\it i.e.}
$\sigma(Q=nk)=0$, which is a consequence of the screening by the
dynamical charge $k$ fermions.

There are some interesting special cases worth mentioning.
First, consider the case with $\theta=\pm\pi$. Then, because
$\ket{\theta}$ and $\ket{\theta\mp 2\pi}$ are degenerate, we have
$\sigma(Q=\mp 1)=0$, which means that strings with the unit flux
can become tensionless. This is probably not too surprising.
From (\ref{Sint}), we see that $\theta=\pm\pi$ can be interpreted as a flux
with $Q=\pm 1/2$. By adding $Q=\mp 1$, we end up with a flux with $Q=\mp 1/2$,
which is obtained by the charge conjugation from the original configuration.

A possibly more surprising case is the $M_0\ra 0$ limit. The formula
(\ref{energydensity}) implies that the tension $\sigma(Q)$ vanishes for
any $Q$ in the $M_0\ra 0$ limit.\cite{Coleman:1975pw,Gross:1995bp,Armoni:1997ki,Armoni:1999xw}
This is a consequence of the fact that the $\theta$-dependence of
the energy density disappears in the massless limit because of the
anomaly relation (\ref{anomaly}). However, if one tries to
understand the vanishing of the $Q$-string tension intuitively as a
screening phenomenon, this looks very strange, because the charge of any
combination of the charge $k$ fermions belongs to $k\Z$ and it doesn't
look possible to completely screen a charge that doesn't belong to $k\Z$.
In our case, the $Q$-string state $\ket{\theta+2\pi Q}$ becomes one of the
ground states in the $M_0\ra 0$ limit. As discussed in section \ref{SSB},
this state is an eigenstate of the operator $\Pi_A$, which contains the
electric flux.
The energy contribution from the eigenvalue of $\Pi_A$ is diminished
(in the $M_0=0$ case) by a constant shift of the scalar field, as the operator
$\Pi_A$ appears in the Hamiltonian in the combination
$(2\pi\Pi_A-kN_f\varphi)^2$ (see (\ref{Hami}) for the $N_f=1$ case).
However, unlike the usual screening phenomenon, the $Q$-string state does not
loose the information of the flux $Q$ in the process of making $M_0\ra 0$,
although it becomes completely tensionless. There is a conserved $\Z_k$ charge
associated with the operator $\wh U'$ defined in (\ref{clockshift})
that characterizes the $Q$-string state.

\section{Application to string theory}
\label{AppString}

In this section, we propose a way to realize 2 dim QED
with $k=2$ in string theory.\footnote{
A string theory realization of 2 dim QED with $k=1$ using
a D1-D9-\AD9 system was studied in \cite{Sugimoto:2004mh}. }
Many of the properties of 2 dim QED studied in section \ref{vacuum}
have natural interpretations in string theory
and the results in section \ref{MassDeform} are applied
to give a new non-perturbative calculation on the brane dynamics.

\subsection{\Om1-\AD1 system} 
\label{o1d1}

Consider a system of $n$ coincident \AD1-branes (anti-D1-branes) on top
of an \Om1-plane (a negative tension orientifold 1-plane).
The low energy effective theory realized on the \AD1-brane world-sheet
is a 2 dim non-supersymmetric $SO(2n)$ gauge theory with
the following massless fields:
\begin{eqnarray}
 \begin{array}{c|ccc}
  &SO(2n)& SO(1,1)&SO(8) \\
\hline
A_\mu & {\rm adj}&2& 1\\
\Phi_I & {\rm adj}&1&8_v\\
\lambda_+^i& {\rm sym}&1_+&8_+\\
\lambda_-^i& {\rm sym}&1_-&8_-\\
 \end{array}
\label{fields1}
\end{eqnarray}
Here, $A_\mu$ ($\mu=0,1$), $\Phi_I$ ($I=1,\cdots,8$)
and $\lambda_\pm^i$  ($i=1,\cdots,8$) are the gauge field,
scalar fields and fermions, respectively.
$SO(2n)$ is the gauge group, $SO(1,1)$ is the 2 dim Lorentz group
and $SO(8)$ is the global symmetry associated to the rotation
in the transverse 8 dim space. The labels ``${\rm adj}$'' and
 ``${\rm sym}$'' refer to the adjoint and rank 2 symmetric tensor
representations of the gauge group, respectively.\footnote{Here,
``sym'' representation is a $n(2n+1)$ dimensional reducible representation
and it can be decomposed to a singlet and rank 2 traceless symmetric tensor
representations.}
$1_\pm$ denotes the positive/negative chirality Majorana Weyl spinor
of $SO(1,1)$. $8_v$ and $8_\pm$ are the vector and positive/negative chirality
spinor representations of $SO(8)$, respectively. 
The field content (\ref{fields1}) is obtained by replacing the adjoint
fermions with that of the rank 2 symmetric tensor representation
in the 2 dim supersymmetric gauge theory realized in
the \Om1-D1 system,\cite{Sugimoto:1999tx}
which is given by the dimensional reduction of
the 10 dim $\cN=1$ supersymmetric $SO(2n)$ Yang-Mills theory.

The gauge coupling $e$ of this 2 dim gauge theory is related to the
string coupling $g_s$ and the string length $l_s=\sqrt{\alpha'}$ by
\begin{eqnarray}
 e^2=\frac{g_s}{2\pi\alpha'}\ .
\label{egs}
\end{eqnarray}
We take the field theory limit  $\alpha'\ra 0$ and $g_s\ra 0$
with $e$ kept fixed, so that stringy massive excitations become
infinitely heavy and interactions with closed string fields including
the gravitational interaction decouple.\footnote{This limit may cause some
divergences in the effective field theory. For example, the one-loop
analysis suggests that a tachyonic mass term for the scalar fields
$\Phi_I$ will be generated and the mass scale will diverge
in the field theory limit.
In renormalizable quantum field theory, such divergences can be
canceled by introducing counter terms and setting the renormalization
conditions to make physical quantities finite. However, it is not
clear whether such counter terms can always be introduced in string theory.
To avoid this problem, we actually keep $l_s$ finite, though we assume
 $e^2\ll 1/\alpha'$,
and consider the system as a theory with the cut-off scale $1/l_s$
regularized by string theory.  (See sections \ref{CWsection}
and \ref{Mobius}.)
}

Let us focus on $n=1$. In this case, since $SO(2)$ is equivalent
to $U(1)$, the theory (\ref{fields1}) becomes a $U(1)$ gauge theory.
The massless fields can be written as
\begin{eqnarray}
 A_\mu=\mat{,-a_\mu,a_\mu,}\ ,~~
 \Phi_I=\mat{,-\phi_I,\phi_I,}\ ,~~
\lambda^i_\pm=\mat{\lambda_\pm^{(0)i}+\lambda_\pm^{(1)i},\lambda_\pm^{(2)i},
\lambda_\pm^{(2)i},\lambda_\pm^{(0)i}-\lambda_\pm^{(1)i}} \ .
\end{eqnarray}
Then, $a_\mu$ is the $U(1)$ gauge field, $\phi_I$ are neutral real
scalar fields, $\lambda_\pm^{(0)i}$ are the neutral Majorana-Weyl
fermions and the complex Weyl fermions defined by
\begin{eqnarray}
\psi^i_L\equiv \lambda^{(1)i}_-+i\lambda^{(2)i}_-\ ,
~~~
\psi^i_R\equiv \lambda^{(1)i}_++i\lambda^{(2)i}_+
\end{eqnarray}
are the charge 2 fermions.
The neutral fermions $\lambda_\pm^{(0)i}$ do not interact with
other fields in the low energy effective action
(in the $\alpha'\ra 0$ limit) and will be neglected
in what follows.
Then, the table (\ref{fields1}) for $n=1$ becomes
\begin{eqnarray}
\begin{array}{l|ccc}
&U(1)~ {\rm charge}& SO(1,1)&SO(8) \\
\hline
a_\mu & 0&2& 1\\
\phi_I & 0&1&8_v\\
\psi^i_R& 2&1_+&8_+\\
\psi^i_L& 2&1_-&8_-\\
 \end{array}
\label{fields2}
\end{eqnarray}
This is almost like a 2 dim QED with $N_f=8$ and $k=2$,
\footnote{
It is possible to introduce an external (infinitely heavy) charge
by putting one end point of a fundamental string on the \AD1-brane,
which has the unit charge. Since $\psi_\pm^i$ have twice of
this charge, we obtain $k=2$.}
but couples with 8 massless scalar fields $\phi_I$. Though
$\phi_I$ do not have gauge interaction, they couple with the fermions
through the Yukawa interaction
\begin{eqnarray}
S_{\rm Yukawa}=\int d^2 x\,\left(
y \Gamma^I_{ij}\, \phi_I\psi_R^{i\dag}\psi_L^j+{\rm h.c.}\right)
\ ,
\label{Yukawa}
\end{eqnarray}
where $y$ is a constant and $(\Gamma^I_{ij})$ is the invariant tensor of
the $8_v\otimes 8_+\otimes 8_-$ representation of $SO(8)$.
(See  Appendix \ref{gamma} for an explicit form.)
Because of this Yukawa interaction, the $SU(N_f)_L\times SU(N_f)_R$
chiral symmetry of 2 dim massless QED is explicitly broken and only
the $SO(8)$ symmetry is manifest.
In the string theory construction, the Yukawa coupling $y$ is not an
independent parameter. When the scalar fields $\phi_I$ are canonically
normalized, $y$ is given as the gauge coupling $e$ times some numerical
constant.

The partition function of the full system is given as
\begin{eqnarray}
Z_{\rm Full}
&=&\int \cD\phi\, e^{i\int d^2x \half(\del_\mu\phi_I)^2}Z_{\rm QED}[\phi_I]\ ,
\nn\\
Z_{\rm QED}[\phi_I]&\equiv&\int\cD a\cD\psi\cD\ol\psi
\, e^{iS_{\rm QED}[a_\mu,\psi^i]+iS_{\rm Yukawa}[\phi_I,\psi^i]}\ .
\end{eqnarray}
In the following, we will focus only on $Z_{\rm QED}[\phi_I]$ and
treat $\phi_I$ as parameters of the system. Furthermore, we assume that
$\phi_I$ are all constant, in which case
\begin{eqnarray}
V_{\rm eff}(\phi_I)\equiv -\frac{i}{V_2}\log Z_{\rm QED}[\phi_I]\ ,
\label{Veff}
\end{eqnarray}
where $V_2\equiv\int d^2 x$ is the volume of 2 dim space-time, is
interpreted as the effective potential for the scalar fields $\phi_I$
that correspond to the position of the \AD1-brane in the transverse
8 dim space.
Using the $SO(8)$ symmetry, we can set $\phi_I=0$ ($I=1,\cdots,7$) and assume
that only $\phi_8\equiv\phi$ can take a non-zero value without loss of generality.
According to (\ref{8gamma}), $\Gamma^8$ is the $8\times 8$ unit matrix
and (\ref{Yukawa}) is simplified as
\begin{eqnarray}
S_{\rm Yukawa}=\int d^2 x\,\left(
y\phi\,\psi_{Ri}^{\dag}\psi_L^i+{\rm h.c.}\right)\ .
\label{Yukawa2}
\end{eqnarray}
This is nothing but the mass term for the fermions considered in section
\ref{MassDeform} with the identification $M_0=y\phi$.
When $\phi$ is non-zero, the \AD1-brane is separated from
the \Om1-plane. Since the fermions $\psi^i_L$ and $\psi^i_R$
are created by the open strings that hung on \Om1-plane
with end points attached on the \AD1-brane, this mass $M_0$
is interpreted as twice the string tension $1/(2\pi\alpha')$
times the distance $Y$ between the \AD1-brane and the \Om1-plane:
\begin{eqnarray}
 M_0=y\phi=\frac{Y}{\pi\alpha'}\ .
\label{M0Y}
\end{eqnarray}

We can also introduce a non-trivial $\theta$ angle in our system.
The parameter $\theta$ is identified as the value of RR 0-form field
$C_0$, which is normalized to be a $2\pi$ periodic scalar field.
However, because the $\Z_2$ orientifold action maps $C_0$ to $-C_0$
($\mod 2\pi$), the allowed value on top of the
\Om1-plane is either $C_0=0$ or $C_0=\pi$ ($\mod 2\pi$).

\subsection{$(Q,-1)$-strings and the short distance potential}

Since the electric flux on the \AD1-brane is interpreted as
the fundamental string, $Q$-strings considered in section
\ref{StrTen} should be interpreted as a bound state
of $Q$ fundamental strings and the \AD1-brane, which is called
a $(Q,-1)$-string.\footnote{In general,
a bound state of $p$ fundamental strings and $q$ D1-branes
is called a $(p,q)$-string.} Note that the $\Z_2$ orientifold action
flips the orientation of the fundamental strings, {\it i.e.} it maps
a fundamental string (F-string) to an anti-fundamental string ($\ol{\rm F}$-string)
in the covering space (the space before modding out the space
by the $\Z_2$ orientifold action).
Therefore, in the covering space, we have a pair of $(Q,-1)$-string and
 $(-Q,-1)$-string that are mapped to each other by the orientifold
action, and hence there is no net F-string charge.

\begin{figure}
\begin{center}
\begin{picture}(370,60)(0,0)
\includegraphics[scale=0.7,bb=0 0 508 69]{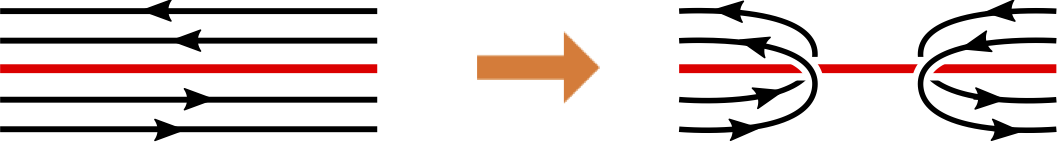}
\put(-371,27){\makebox(0,0){\Om1}}
\put(-330,55){\makebox(0,0){F-string}}
\put(-330,-7){\makebox(0,0){$\ol{\rm F}$-string}}
\end{picture}
\vspace*{3mm}
\caption{
Reconnection of two F-string - $\ol{\rm F}$-string pairs.
}
\label{reconn}
\end{center}
\end{figure}

As we have seen in section \ref{StrTen}, $Q$-string and $(Q+k)$-string
are identified, because $k$ unit of electric flux can be screened
by the charge $k$ fermions. In our $k=2$ case, it implies that
the only non-trivial one is the $(1,-1)$-string.
In terms of the pair of $(Q,-1)$-string and $(-Q,-1)$-string
in the covering space, this can be understood from the fact
that an even number of F-string - $\ol{\rm F}$-string
pairs can be annihilated by reconnection. (See Figure \ref{reconn}.)
On the other hand,
a single  F-string - $\ol{\rm F}$-string pair cannot be annihilated.
In order for the single F-string - $\ol{\rm F}$-string pair
to be reconnected, the F-string has to go through the orientifold
plane to be consistent with the orientifold action.  However,
the \Om p-plane does not allow such configurations.\footnote{
It is possible to have such configurations for the \Omt p-plane,
on which F-strings can end.
}

The $Q$-string tension (\ref{tension}) corresponds to the
difference between the $(Q,-1)$-string tension, denoted as
$T_{(Q,-1)}$, and the \AD1-brane tension $T_{(0,-1)}$.
Using
(\ref{ftheta}), (\ref{energydensity}), (\ref{egs}) and
(\ref{M0Y}), the $(Q,-1)$-string tension is obtained as
\begin{eqnarray}
T_{(Q,-1)}
=T_0-C_Q\frac{g_s}{\alpha'}
\left(\frac{Y^2}{g_s\alpha'}\right)^{\frac{8}{9}}
\ ,
\label{TQ}
\end{eqnarray}
where $T_0$ is a constant that does not depend on $\theta$, $Q$ and $Y$,
and $C_Q$ is given by
\begin{eqnarray}
C_Q=
\frac{18}{\pi^3}
\left(
\frac{\exp\left(\gamma\right)}{2}
\cos\left(\frac{1}{8}\ol{\left(\theta/2+\pi Q\right)}
\right)
\right)^{\frac{16}{9}}
\ .
\label{CQ}
\end{eqnarray}
This expression is valid when $M_0\ll e$, which is equivalent to
$Y^2\ll g_s\alpha'$.
Note that the second term in (\ref{TQ}) contains the string coupling
with a fractional power as $g_s^{\frac{1}{9}}$.
This is a non-perturbative prediction at
short distances.

When $\theta=\pm\pi$, (\ref{CQ}) implies
\begin{eqnarray}
C_Q=
\frac{18}{\pi^3}
\left(
\frac{\exp\left(\gamma\right)}{2}
\cos\left(\frac{\pi}{16}\right)
\right)^{\frac{16}{9}}
\ ,
\end{eqnarray}
which is independent of $Q$. Note that
$T_{(\mp 1,-1)}=T_{(0,-1)}$ for $\theta=\pm\pi$  follows from the fact
that $\sigma(Q=\mp 1)$ vanishes when $\theta=\pm\pi$ as
discussed in section \ref{StrTen}.
This is true even when the \AD1-brane is far away from the \Om1-plane,
in which case the well-known formula for the $(p,q)$-string tension
\begin{eqnarray}
 T_{(p,q)}=\frac{1}{2\pi\alpha'}
\sqrt{\left(p-q\frac{\theta}{2\pi}\right)^2+\frac{q^2}{g_s^2}}
\label{Tpq}
\end{eqnarray}
with $\theta=\pm\pi$ implies the same conclusion.

For $\theta=0$, (\ref{CQ}) implies
\begin{eqnarray}
C_Q=
\frac{18}{\pi^3}\left(\frac{\exp\left(\gamma\right)}{2}
\right)^{\frac{16}{9}}
\times\left\{
\begin{array}{cc}
1&(Q={\rm even})\\
\cos^{\frac{16}{9}}\left(\frac{\pi}{8}\right)
&(Q={\rm odd})
\end{array}
\right.
\ .
\end{eqnarray}
In particular, we obtain
\begin{eqnarray}
 T_{(1,-1)}-T_{(0,-1)}
=C\,
\frac{g_s}{\alpha'}
\left(\frac{Y^2}{g_s \alpha'}\right)^{\frac{8}{9}}\ ,
\label{shortT}
\end{eqnarray}
where
\begin{eqnarray}
 C\equiv \frac{18}{\pi^3}
\left(\frac{\exp\left(\gamma\right)}{2}\right)^{\frac{16}{9}}
\left(
1-\cos^{\frac{16}{9}}\left(\frac{\pi}{8}\right)
\right)
\end{eqnarray}
is a positive constant.
This result is in contrast to the behavior at long distances
($\alpha'\ll Y^2$) obtained from (\ref{Tpq}):
\begin{eqnarray}
 T_{(1,-1)}-T_{(0,-1)}=\frac{1}{2\pi\alpha' g_s}
\left(\sqrt{1+g_s^2}-1\right)
\simeq \frac{g_s}{4\pi\alpha'}\ ,
\end{eqnarray}
where we have used $g_s\ll 1$.

As noted in the last paragraph of section \ref{StrTen},
the right hand side of (\ref{shortT}) vanishes in the $Y\ra 0$ limit.
Therefore, when the $(1,-1)$-string is placed on top of the \Om1-plane,
the energy contribution from the fundamental string completely
disappears and the tension become the same as the \AD1-brane
without the electric flux. This phenomenon should not be interpreted
as the annihilation of the F-string - $\ol{\rm F}$-string pair,
as emphasized in section \ref{StrTen}.

The short distance potential between the $(Q,-1)$-string
and the \Om1-plane is given by (\ref{TQ}).
This formula is obtained by evaluating the energy density
$\cE$ in (\ref{energydensity}) as a function of $Y$ related
to the fermion mass $M_0$ by (\ref{M0Y}). Note that
this agrees with the effective potential $V_{\rm eff}(Y)$
defined in (\ref{Veff}) up to $Y$ independent constant terms, as
\begin{eqnarray}
Y\frac{\del V_{\rm eff}(Y)}{\del Y}
=-\frac{i}{V_2}\frac{M_0\del_{M_0}Z_{\rm QED}}{Z_{\rm QED}}
=M_0\VEV{\ol\psi_i\psi^i}=Y\frac{\del\cE(Y)}{\del Y}\ ,
\end{eqnarray}
where we have used (\ref{EM0rel}) in the last step.

The expression (\ref{TQ}) shows that there is a repulsive force between
the $(Q,-1)$-string and the \Om1-plane. It is interesting to note that
this force is proportional to the chiral condensate in 2 dim QED
as (\ref{EM0rel}) and (\ref{M0Y}) implies
\begin{eqnarray}
 \VEV{\ol\psi_j\psi^j}=\pi\alpha'\frac{\del\cE(Y)}{\del Y}\ .
\label{cond-force}
\end{eqnarray}

\subsection{Coleman-Weinberg potential}
\label{CWsection}

Within the field theory limit, the long distance potential between the
\Om1-plane and the \AD1-brane can be calculated using a Coleman-Weinberg
potential \cite{Coleman:1973jx}. Indeed, in our system, perturbation
theory (with respect to the gauge coupling $e$) can be trusted when the
vacuum expectation value
of the scalar field $\phi$ is large and satisfies $e\ll M_0=y\phi$,
which corresponds to $g_s\alpha'\ll Y^2$.
In order for the field theory description to be valid,
$e^2\ll 1/\alpha'$ and $M_0^2\ll 1/\alpha'$ have to be satisfied,
which is equivalent to $g_s\ll 1$ and $Y^2\ll\alpha'$.

The one-loop Coleman-Weinberg potential in 2 dim is written in general as
\begin{eqnarray}
 V_{\rm eff}(\phi)=V_{\rm tree}(\phi)
+\frac{1}{2}\int \frac{d^2 k}{(2\pi)^2}
\left[
\sum_{b:{\rm boson}}\log\left(\frac{k^2+m_b^2(\phi)}{k^2}\right)
-\sum_{f:{\rm fermion}}\log\left(\frac{k^2+m_f^2(\phi)}{k^2}\right)
\right]\ ,
\label{CW}
\end{eqnarray}
where $V_{\rm tree}(\phi)$ is the tree level potential, 
$k$ is the momentum in the Euclidean space, and $m_b$ and $m_f$
are the mass for the bosonic field $b$ and the fermionic field $f$, respectively.

The integral can be evaluated as follows:
\begin{eqnarray}
&& \half\int \frac{d^2 k}{(2\pi)^2}
\log\left(\frac{k^2+m^2}{k^2}\right) = \frac{1}{4\pi}\int_0^{\Lambda} d k\,k
\log\left(\frac{k^2+m^2}{k^2}\right)
\nn\\
&=&
\frac{1}{8\pi}\left[
m^2\log\left(\frac{\Lambda^2}{m^2}\right)
+(m^2+\Lambda^2)\log\left(1+\frac{m^2}{\Lambda^2}\right)
\right] \simeq
\frac{m^2}{8\pi}\left[
\log\left(\frac{\Lambda^2}{m^2}\right)+1
\right]\ ,
\label{int1}
\end{eqnarray}
where $\Lambda$ is the cut-off scale.
In the last step, we neglected the terms that vanishes in the
$\Lambda\ra\infty$ limit.

In our system (\ref{fields2}),
only the fermions $\psi^i$ have $\phi$ dependent mass via
the Yukawa term (\ref{Yukawa}) and hence we obtain
\begin{eqnarray}
V_{\rm eff}(Y) \simeq -\frac{M_0^2}{8\pi}\left[
\log\left(\frac{\Lambda^2}{M_0^2}\right)+1\right]
=-\frac{Y^2}{8\pi^3\alpha'^2}\left[
\log\left(\frac{\pi^2\alpha'^2\Lambda^2}{Y^2}\right)+1
\right]
 \ ,
\label{VeffCW}
\end{eqnarray}
which again shows a repulsive force.

\subsection{The potential from the M\"obius strip amplitude}
\label{Mobius}

For completeness, let us review the calculation of the potential
between an orientifold plane and an anti D-brane using string theory.
This calculation takes into account the stringy tower and it is
valid at long distances, where $Y^2$ is of order $\alpha'$ or larger
and the effective field theory description breaks down.

\begin{figure}
\centerline{\includegraphics[scale=0.8,bb=0 0 260 160]{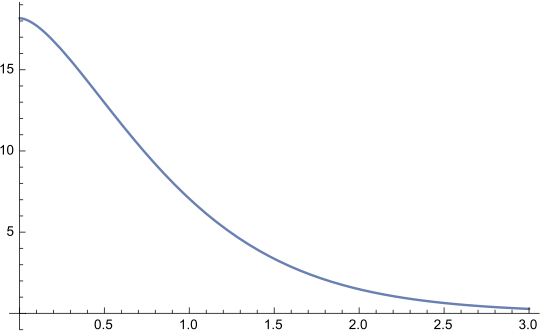}
\put(10,5){\makebox(0,0){$Z$}}
\put(-200,140){\makebox(0,0){$V(Z)$}}
}
\vspace*{3mm}
\caption{
Potential for the \Om1-\AD1 system:
$V(Z)=\int_0^{\infty}\frac{dt}{2t^2}e^{-tZ^2}F_O(t)$}
\label{VO}
\end{figure}

Consider a system with an \Opm p-plane and a \AD p-brane
placed in parallel with distance $Y$. The potential between
the \Opm p-plane and \AD p-brane is given by
an open string vacuum amplitude corresponding to
the world-sheet with topology of the M\"obius strip at one-loop level.
The open string one-loop amplitude can be interpreted
as a closed string tree level amplitude and the potential is
generated by the exchange of graviton, RR fields and so on.
The explicit calculation of the M\"obius strip amplitude
for a system with parallel \Opm p-plane and \AD p-brane
was carried out in \cite{Uranga:1999ib}, and the result is
\begin{eqnarray}
V^{\mbox{\tiny \Opm{p}-\AD{p}}}_{\rm eff}(Y)&=&
\mp \int^\infty_0\frac{dt}{2t}(8\pi^2\alpha't)^{-\frac{p+1}{2}}
e^{-t\frac{(2Y)^2}{2\pi\alpha'}}F_O(t)\ ,
\label{Vtotal}
\end{eqnarray}
where
\begin{eqnarray}
F_O(t)\equiv \frac{Z^0_1(2it)^4Z^1_0(2it)^4}{\eta(2it)^8Z^0_0(2it)^4}\ .
\label{FOdef}
\end{eqnarray}
The functions $\eta(it)$ and $Z^\alpha_\beta(it)$ are defined as
\begin{eqnarray}
\eta(it)&=&
q^{1/24}\prod_{m=1}^\infty(1-q^{m})\ ,\\
 Z^0_0(it)&=&
q^{-1/24}\prod_{m=1}^\infty(1+q^{m-1/2})^2\ ,\\
 Z^0_1(it)&=&
q^{-1/24}\prod_{m=1}^\infty(1-q^{m-1/2})^2\ ,\\
 Z^1_0(it)&=&
2q^{1/12}\prod_{m=1}^\infty(1+q^{m})^2\ ,
\end{eqnarray}
where $q=e^{-2\pi t}$.
They satisfy
\begin{eqnarray}
\eta(it)=t^{-1/2}\eta(i/t)\ ,~~~
 Z^\alpha_\beta(it)^4=Z^\beta_\alpha(i/t)^4\ ,
\label{modular}
\end{eqnarray}
and
\begin{eqnarray}
Z^0_0(it)^4-Z^0_1(it)^4-Z^1_0(it)^4=0\ .
\label{abstruse}
\end{eqnarray}
One can check that
$F_O(t)$ is a positive monotonically increasing function that interpolates
$F_O(t=0)=0$ and $F_O(t=\infty)=16$. For small $t$, it behaves as
$F_O(t)\simeq 2^8 t^4$ and hence the integral over $t$ in (\ref{Vtotal})
is convergent when $p<7$.
The potential is positive (negative) monotonically decreasing (increasing)
function of $Y$ for O$p^-$-\AD$p$ system (O$p^+$-\AD$p$ system, respectively).
In particular, for the \Om1-\AD1 system we find a repulsion.
See Figure \ref{VO} for its shape for the \Om1-\AD1 system.

A few comments are in order.
The function $F_O$ can be written as
\begin{eqnarray}
F_{O}(t)= \sum_{b:{\rm boson}}e^{-t(2\pi\alpha')m_b^2}
- \sum_{f:{\rm fermion}}e^{-t(2\pi\alpha')m_f^2}\ ,
\end{eqnarray}
and the expression (\ref{Vtotal}) is interpreted as a sum of
the contributions from all the fields in the spectrum with
\begin{eqnarray}
m_b^2(\phi)=m_b^2+\left(\frac{2Y}{2\pi\alpha'}\right)^2\ ,~~~
m_f^2(\phi)=m_f^2+\left(\frac{2Y}{2\pi\alpha'}\right)^2
\end{eqnarray}
in the Coleman-Weinberg potential ((\ref{CW}) for the $p=1$ case).

In particular, $F_O(t=\infty)=16$ corresponds to the contribution
from the 16 complex massless Weyl fermions ($\psi_R^i$ and
$\psi_L^i$ in our case).
The contributions from these massless modes are dominant at short
distances and the potential smoothly approaches the Coleman-Weinberg
potential ((\ref{VeffCW}) for $p=1$) obtained in field theory.

One can easily show that the potential (\ref{Vtotal}) behaves as
\begin{eqnarray}
V^{\mbox{\tiny \Opm{p}-\AD{p}}}_{\rm eff}(Y)\simeq
\mp 2^p(8\pi^2\alpha')^{-\frac{p+1}{2}}
\left(\frac{2\pi\alpha'}{Y^2}\right)^{\frac{7-p}{2}}
\Gamma\left(\frac{7-p}{2}\right)
\end{eqnarray}
for $\alpha'\ll Y^2$. This is interpreted as the potential
generated by the exchange of supergravity fields.
It is known that there is no force between an \Opm p-plane
and a D$p$-brane due to supersymmetry. When we replace
the D$p$-brane with the \AD p-brane, the sign of the RR charge
is flipped from positive to negative and the force balance
is broken. Because \Om p-plane (\Op p-plane) has negative
(positive) RR-charge, the \AD p-brane is repelled from 
(attracted to) the \Om p-plane (\Op p-plane, respectively).

\section{Summary and Discussion}
\label{conclusions}

In this paper, we first considered an interesting variant of 2 dim
QED with $N_f$ flavours of fermions with charge $k$. We found a rich
vacuum structure of $k$ vacua, parametrized by
the vacuum expectation value of $\det(\psi_{Rj}^\dag\psi_L^i)$
as an order parameter. We also studied the $Q$-string tension
including the quark mass.

In section \ref{AppString}, we used the results of the first part to
learn about non-perturbative string dynamics. The vacuum energy
as a function of the quark mass can be interpreted as
the potential between the \Om1-plane and the \AD1-brane.
In addition, we could also use the energy density for the $Q$-string
to learn about the dynamics of the $(Q,-1)$-string placed
near the \Om1-plane.
We found that the $(Q,-1)$-string tension (\ref{TQ})
admits a non-perturbative dependence on the
string coupling of the form $g_s^{\frac{1}{9}}$.

One might think that the existence of $k$ vacua would imply domain walls
that interpolate between these vacua.
If such an object exists, it would make the $Q$-string states unstable
by the creation of domain wall - anti domain wall pairs.
However, it is not allowed in our case. As we have seen
in section \ref{SSB}, the vacuum is parametrized by $\theta$,
which is an eigenvalue of the operator $2\pi\Pi_A$.
Because of the Gauss law equation (\ref{GLaw}), we cannot have
an object that interpolates different values of $\theta$
in $x^1\ra-\infty$ and $x^1\ra+\infty$.\footnote{
It is possible to set a boundary condition that the scalar field
$\varphi$ in the bosonized description (\ref{Sboson-multi})
approaches different values at the boundaries $x^1\ra-\infty$
and $x^1\ra+\infty$. However, in that case, the electric flux $F_{01}$
will be inevitably induced (at least) at one of the boundaries, due to
the Gauss law equation (\ref{GLaw}). Therefore, it does not describe a
domain wall that connects two vacua related by the discrete axial
symmetry.}
In terms of string theory, if one wants to consider a configuration
with a $(Q,-1)$-string at $x^1\ra-\infty$ and 
 $(Q+1,-1)$-string at $x^1\ra+\infty$, there must be a fundamental
string attached on it. Since the other end point of the fundamental
string should escape to spatial infinity, it is infinitely heavy.

Part of the discussion of the previous sections could be used in
higher dimensional cases. 
Consider a system of \Om p-plane and a \AD p-brane. Similar to
our 2 dim case, the low energy theory of this system is described by a
$(p+1)$ dim QED with charge 2 fermions coupled with neutral scalar
fields via Yukawa coupling.
The expressions analogous to (\ref{cond-force}) would also hold in higher
dimensional cases. In general, we expect that the vacuum expectation
value of the fermion bilinear operator is non-zero when
the fermion mass is turned on and hence the relation
(\ref{cond-force}) implies a force between them.
This is a typical relation that connects a vacuum expectation value
of an operator in quantum field theory and a quantity that characterizes
the brane dynamics. It would be interesting to explore the
generalization of such relations and find more fruitful interplay
between quantum field theory and string theory.

\subsection*{Acknowledgements}

We thank Sinya Aoki, Aleksey Cherman, Satoshi Iso, Zohar Komargodski,
Erich Poppitz, Satoshi Yamaguchi, Piljin Yi and Yang Zhou for
discussions and helpful comments.
We also appreciate useful discussions during the workshops
``Simons Summer Workshop 2017'' held at the Simons Center for Geometry and
Physics at Stony Brook and
``New Frontiers in String Theory 2018'' Ref. YITP-T-18-04
held at the Yukawa Institute for Theoretical Physics, Kyoto University.
SS is also grateful to the participants of the workshops
``East Asia Joint Workshop on Fields and Strings 2018'' held
at KIAS and ``KEK Theory Workshop 2018'' held at KEK, in which
he made presentations on this work and got
a lot of valuable feedback.
The work of AA has been supported by STFC grant ST/P00055X/1.
The work of SS has been supported by JSPS KAKENHI (Grant-in-Aid for Scientific
Research (C)) Grant Number JP16K05324.

 
\appendix

\section{The explicit form of $\Gamma^I_{ij}$}
\label{gamma}

In this section,
we list an explicit form of $\Gamma^I_{ij}$ used in (\ref{Yukawa}).
The $8\times 8$ matrices $\Gamma^I=(\Gamma^I_{ij})$
are related to the $SO(8)$ gamma matrices $\wh\Gamma^I$ as
\begin{eqnarray}
\wh\Gamma^I=\mat{,\Gamma^I,{}^t\Gamma^I,}\ ,
\end{eqnarray}
and an explicit form is given as
\begin{eqnarray}
\begin{array}{llll}
\Gamma^1=\epsilon\otimes\epsilon\otimes\epsilon\ ,&
\Gamma^2=1_2\otimes\tau_1\otimes\epsilon\ ,&
\Gamma^3=1_2\otimes\tau_3\otimes\epsilon\ ,&
\Gamma^4=\tau_1\otimes\epsilon\otimes 1_2\ ,
\\
\Gamma^5=\tau_3\otimes\epsilon\otimes 1_2\ ,&
\Gamma^6=\epsilon\otimes 1_2\otimes\tau_1\ ,&
\Gamma^7=\epsilon\otimes 1_2\otimes \tau_3\ ,&
\Gamma^8=1_2\otimes 1_2\otimes 1_2\ ,
\end{array}
\label{8gamma}
\end{eqnarray}
where $\epsilon\equiv i\tau_2$ 
and $\tau_j$ ($j=1,2,3$) are the Pauli matrices.\footnote{
See p.288 of \cite{Green:1987sp}.}

\section{Abelian bosonization, chiral condensate and energy density}

In this section, we outline the calculation of the chiral condensate
and the vacuum energy density in the strong coupling regime
for the $k=1$ case, using the Abelian bosonization.

\subsection{Abelian bosonization}

Here, we briefly review the Abelian bosonization.\footnote{
See, e.g., \cite{Frishman:1992mr} for a review}
The idea is to bosonize the $N_f$ flavours of fermions one by one
and map the system to a theory with
$N_f$ real scalar fields. This description has a disadvantage
that the $SU(N_f)_L\times SU(N_f)_R$ symmetry is not manifest.
But, it is often used because it is simpler and useful for calculations.

Let us consider 2 dim QED with $N_f$ massive Dirac fermions of charge $k=1$.
The bosonization rules are \cite{Coleman:1974bu}

\begin{eqnarray}
\ol\psi_i\gamma^\mu\psi^i
&=&\frac{1}{\sqrt{\pi}}\epsilon^{\mu\nu}\del_\nu\varphi_i\ ,
\label{current}
\\
\ol\psi_i\psi^i
&=&-c\, \mu \cN_\mu\cos(2\sqrt{\pi}\varphi_i)\ ,
\label{mass}
\end{eqnarray}
where the index $i=1,2,\cdots,N_f$ in the left hand side is not summed over,
$\varphi_i$ are canonically normalized real scalar fields and  $c$ is
a constant
\begin{eqnarray}
c\equiv \frac{\exp(\gamma)}{2\pi}\ .
\label{c}
\end{eqnarray}
$\cN_\mu$ denotes the normal ordering with respect to an arbitrary
scale $\mu$. Useful identities (see \cite{Coleman:1974bu}) are
\begin{eqnarray}
\cN_\mu\left[\half\left(\Pi^2+(\del_1\varphi)^2\right)\right]
&=&\cN_{\mu'}\left[\half\left(\Pi^2+(\del_1\varphi)^2\right)\right]
+\frac{1}{8\pi}(\mu'^2-\mu^2)\ ,
\label{renormM1}
\\
\cN_\mu\left[\half m^2\varphi^2 \right]
&=&\cN_{\mu'}\left[\half m^2\varphi^2 \right]
-\frac{m^2}{8\pi}\log\left(\frac{\mu'^2}{\mu^2}\right)\ ,
\label{renmass}
\end{eqnarray}
where $\Pi$ is the canonical momentum operator conjugate to $\varphi$,
and
\begin{eqnarray}
\cN_\mu e^{i\beta\varphi}=\left(\frac{\mu'}{\mu}\right)^{\beta^2/4\pi}
\cN_{\mu'}e^{i\beta\varphi}\ .
\label{ren}
\end{eqnarray}
(\ref{ren}) implies
\begin{eqnarray}
\mu\cN_\mu\cos(2\sqrt{\pi}\varphi)=\mu'\cN_{\mu'}\cos(2\sqrt{\pi}\varphi)\ .
\label{renormM2}
\end{eqnarray}

Then, after integrating out the gauge field, the bosonized Hamiltonian (density)
for the $N_f$ flavour massive Schwinger model is
\begin{eqnarray}
\cH&=&
\cN_\mu\left[
\half\sum_{i=1}^{N_f}\left(\Pi_i^2+(\del_1\varphi_i)^2\right)
+\frac{e^2}{2\pi}\left(\sum_{i=1}^{N_f}\varphi_i\right)^2
-c\, \mu M_0\sum_{i=1}^{N_f}\cos\left(2\sqrt{\pi}\varphi_i
+\frac{\theta}{N_f}
\right)
\right]
\nn\\
&&+\frac{N_f}{8\pi}\left(\mu^2-\frac{e^2}{\pi}\log \mu^2\right)
\ ,
\label{Nfflavor}
\end{eqnarray}
where $M_0$ is the fermion mass.
Note that the Hamiltonian (\ref{Nfflavor}) does not depend on the
renormalization scale $\mu$ because of
the relations (\ref{renormM1}), (\ref{renmass}) and (\ref{renormM2}).
This is the reason that we put the constant term in (\ref{Nfflavor}).

In the Hamiltonian (\ref{Nfflavor}), the $2\pi$ periodicity of $\theta$
is not manifest. But, it is easy to see that the Hamiltonian is
invariant under $2\pi$ shift of $\theta$ together with the transformation
\begin{eqnarray}
2\sqrt{\pi}\varphi_1&\ra&2\sqrt{\pi}\varphi_1-\frac{2\pi}{N_f}+2\pi\ ,
\nn\\
2\sqrt{\pi}\varphi_i&\ra&2\sqrt{\pi}\varphi_i-\frac{2\pi}{N_f}
\ ,~~~(i=2,3,\cdots,N_f)\ .
\label{2pi}
\end{eqnarray}
Therefore, physics is invariant under the $2\pi$ shift of $\theta$.

\subsection{Chiral condensate and energy density}
\label{AppCC}

We are interested in the cases with $e^2\gg \mu M_0$ and try to integrate
out the heavy combination $\sum_{i=1}^{N_f}\varphi_i$ first.
To do this, it will be convenient to introduce a matrix notation:
\begin{eqnarray}
\Phi\equiv\diag(\varphi_1,\varphi_2,\cdots,\varphi_{N_f})\ ,
\end{eqnarray}
and
\begin{eqnarray}
h&\equiv&\tr(T^0\Phi)\ ,
\\
l_i&\equiv&\tr\left(T^i \Phi\right)
\ ,~~~(i=1,2,\cdots,N_f-1)\ ,
\end{eqnarray}
where $T^0\equiv\frac{1}{\sqrt{N_f}}1_{N_f}$,
and
 $T^i$ $(i=1,2,\cdots,N_f-1)$ are the generators
of the Cartan subalgebra of $su(N_f)$ (diagonal
traceless Hermitian matrices) normalized as
\begin{eqnarray}
\tr(T^iT^j)=\delta^{ij}\ .
\end{eqnarray}
Then, $\Phi$ can be expanded as
\begin{eqnarray}
 \Phi=hT^0+\sum_{i=1}^{N_f-1}l_iT^i\ ,
\end{eqnarray}
and the Hamiltonian becomes
\begin{eqnarray}
\cH
&=&
\cN_\mu\left[
\half\tr\left(\Pi_\Phi^2+(\del_1\Phi)^2\right)
+\frac{e^2}{2\pi}\left(\tr\Phi\right)^2
-c\, \mu M_0\tr\left(\cos\left(2\sqrt{\pi}\Phi
+\frac{\theta}{N_f}
\right)\right)\right]
\nn\\
&&+\frac{N_f}{8\pi}\left(\mu^2-\frac{e^2}{\pi}\log \mu^2\right)
\nn\\
&=&
\cN_\mu\Bigg[\half(\Pi_h^2+(\del_1h)^2)
+\half\sum_{i=1}^{N_f-1}\left(\Pi_{l_i}^2+(\del_1l_i)^2\right)
+\frac{m_h^2}{2}h^2
+\frac{N_f}{8\pi}\left(\mu^2-\frac{e^2}{\pi}\log \mu^2\right)
\nn\\
&&
-\frac{c\, \mu M_0}{2}e^{i\sqrt{\frac{4\pi}{N_f}}h+i\frac{\theta}{N_f}}
\tr\left(e^{i\,2\sqrt{\pi}
\sum_{i=1}^{N_f-1}l_iT^i}
\right)+{\rm h.c.}
\Bigg]\ ,
\label{Ham}
\end{eqnarray}
where we have defined
the mass scale of the heavy component $h$ as
\begin{eqnarray}
m_h^2\equiv\frac{e^2N_f}{\pi} \ .
\end{eqnarray}

For $e^2\gg \mu M_0$, the heavy field $h$ can be treated as a free
massive scalar field of mass $m_h$ and the operator
$\cN_\mu e^{i\sqrt{\frac{4\pi}{N_f}}h}$ in the Hamiltonian (\ref{Ham})
can be replaced with its vacuum expectation value in the low energy
effective theory for the light fields $l_i$.
To evaluate the vacuum expectation value, it is convenient to choose
the scale of the normal ordering to be $m_h$ using the formula
\begin{eqnarray}
\cN_\mu e^{i\sqrt{\frac{4\pi}{N_f}}h}
=\left(\frac{m_h}{\mu}\right)^{\frac{1}{N_f}}
 \cN_{m_h}e^{i\sqrt{\frac{4\pi}{N_f}}h}\ ,
\label{renorm}
\end{eqnarray}
which follows from (\ref{ren}). Using (\ref{renorm}), we obtain
\begin{eqnarray}
 \VEV{\cN_{\mu} e^{i\sqrt{\frac{4\pi}{N_f}} h}}
=\left(\frac{m_h}{\mu}\right)^{\frac{1}{N_f}}\ .
\end{eqnarray}
Similarly, using (\ref{renormM1}), (\ref{renmass}) and
$\VEV{\cN_{m_h}\left[\half\left(\Pi_h^2+(\del_1h)^2+m_h^2h^2\right)\right]}=0$,
 we get
\begin{eqnarray}
\VEV{\cN_\mu\left[\half\left(\Pi_h^2+(\del_1h)^2+m_h^2h^2
\right)\right]}=
\frac{1}{8\pi}\left(m_h^2-\mu^2-m_h^2\log\left(\frac{m_h^2}{\mu^2}\right)
\right)\ .
\end{eqnarray}

Then, the effective Hamiltonian for the light fields $l_i$ is
\begin{eqnarray}
\cH_l
&=&
\cN_\mu\Bigg[\half\sum_{i=1}^{N_f-1}\left(\Pi_{l_i}^2+(\del_1l_i)^2\right)
-\frac{c\, \mu M_0}{2}\left(\frac{m_h}{\mu}\right)^{\frac{1}{N_f}}
e^{i\frac{\theta}{N_f}}
\tr\left(
e^{i\,2\sqrt{\pi}
\sum_{i=1}^{N_f-1}l_iT^i}
\right)+{\rm h.c.}
\Bigg]
\nn\\
&&
+\frac{N_f}{8\pi}\left(\mu^2-\frac{e^2}{\pi}\log \mu^2\right)
+\frac{1}{8\pi}\left(m_h^2-\mu^2-m_h^2\log\left(\frac{m_h^2}{\mu^2}\right)
\right)
\label{Ham2}\\
&=&
\cN_\mu\Bigg[\half\sum_{i=1}^{N_f-1}\left(\Pi_{l_i}^2+(\del_1l_i)^2\right)
-c\, \mu M_0\left(\frac{m_h}{\mu}\right)^{\frac{1}{N_f}}
\cos\left(\frac{\theta}{N_f}\right)
\left(
N_f-2\pi\sum_{i=1}^{N_f-1}l_i^2
\right)+\cO(l_i^3)
\Bigg]
\nn\\
&&
+\frac{N_f-1}{8\pi}\mu^2
+\frac{m_h^2}{8\pi}\left(1-\log m_h^2\right)
\ . 
\label{Hl}
\end{eqnarray}
In the last expression in (\ref{Hl}), we expanded $\cH_l$
with respect to $l_i$ to extract the mass term for $l_i$.
Here, we have assumed that $l_i=0$ is the vacuum
configuration. This is true for $-\pi\le\theta\le\pi$,
but not for general $\theta$. To see the vacuum configuration
for general $\theta$, let us assume
that the vacuum configuration of $l_i$ satisfies
\begin{eqnarray}
\exp\left(i\,2\sqrt{\pi}\sum_{i=1}^{N_f-1}l_iT^i\right)=e^{i\alpha}1_{N_f}\ ,
\label{lvac}
\end{eqnarray}
where $\alpha\in\R$ and $1_{N_f}$ is the unit matrix of size $N_f$,
so that the flavour symmetry is not broken. Because the left hand side
of (\ref{lvac}) is a diagonal element of $SU(N_f)$, $\alpha$ has to be
of the form $\alpha=2\pi\frac{n}{N_f}$ ($n=1,\cdots,N_f$). Then, the potential
term for $l_i$ in the first line of (\ref{Ham2}) is given by a positive constant
times $-\cos((\theta+2\pi n)/N_f)$, which is minimized when $n$ is chosen
so that it satisfies $-\pi\le \theta+2\pi n\le \pi$. Then, expanding
the fields $l_i$ around this configuration is equivalent to (\ref{Hl})
with $\theta$ replaced by
\begin{eqnarray}
 \ol\theta\equiv\theta-2\pi\left[\frac{\theta+\pi}{2\pi}\right]\ ,
\end{eqnarray}
where $[x]$ is the floor function
that gives the greatest integer less than or equal to $x$.
By definition, the value of $\ol\theta$ is restricted to be in the
interval $-\pi\le\ol\theta\le \pi$ and ensures the $2\pi$ periodicity
of the $\theta$ parameter.
Then, the mass scale of the light fields $l_i$ is given by
\begin{eqnarray}
 m_l^2\equiv 4\pi c\, \mu
M_0\left(\frac{m_h}{\mu}\right)^{\frac{1}{N_f}}
\cos\left({\frac{\ol\theta}{N_f}}\right)\ .
\end{eqnarray}

In order to find the ground state, we employ the variational method
used in \cite{Coleman:1974bu}. We first assume that the ground state is
given as $\ket{0;\mu}$, which is the state annihilated by
the annihilation operators defined by the fields $l_i$ with
the mass scale $\mu$ in the Schr\"odinger picture
(see \cite{Coleman:1974bu}) and then find the value of $\mu$
that minimizes the expectation value of the Hamiltonian.
Although we will not try to prove that there is no state
with lower energy, it gives a candidate for the ground state.
In fact, we will show that the expectation value of the fermion
bilinear operator with respect to the state $\ket{0,\mu}$,
denoted as $\VEV{\ol\psi_i\psi^i}_\mu\equiv
\bra{0,\mu}\ol\psi_i\psi^i\ket{0,\mu}$, reproduces
the results in \cite{Hetrick:1995wq,Hetrick:1995yx}.

The expectation value of the Hamiltonian (\ref{Ham2})
with respect to the state $\ket{0,\mu}$ is evaluated as
\begin{eqnarray}
\VEV{\cN_\mu\Bigg[\half\sum_{i=1}^{N_f-1}\left(\Pi_{l_i}^2+(\del_1l_i)^2\right)
\Bigg]+\frac{N_f-1}{8\pi}\mu^2}_\mu
=\frac{N_f-1}{8\pi}\mu^2\ ,
\end{eqnarray}
\begin{eqnarray}
\VEV{\cN_\mu\Bigg[
-\frac{c\, \mu M_0}{2}\left(\frac{m_h}{\mu}\right)^{\frac{1}{N_f}}
e^{i\frac{\theta}{N_f}}
\tr\left(
e^{i\,2\sqrt{\pi}
\sum_{i=1}^{N_f-1}l_iT^i}
\right)+{\rm h.c.}
\Bigg]}_\mu
=-\frac{N_f}{4\pi} m_l^2\ ,
\label{VEVpsibarpsi}
\end{eqnarray}
and
\begin{eqnarray}
\VEV{\cH_l}_\mu=
\frac{N_f-1}{8\pi}\mu^2-\frac{N_f}{4\pi}m_l^2
+\frac{m_h^2}{8\pi}\left(1-\log m_h^2\right)\ .
\end{eqnarray}
Then, it is easy to see that the value of $\mu$
that minimizes this expression satisfies
\begin{eqnarray}
\mu^2= m_l^2\ ,
\label{muml}
\end{eqnarray}
which implies
\begin{eqnarray}
 m_l= \left(4\pi c\, \cos\left(\frac{\ol\theta}{N_f}\right)
 M_0\right)^{\frac{N_f}{N_f+1}}m_h^{\frac{1}{N_f+1}}\ ,
\end{eqnarray}
and
\begin{eqnarray}
\VEV{\cH_l}_{\mu=m_l}=
-\frac{N_f+1}{8\pi}
\left(4\pi
 c\,\cos\left(\frac{\ol\theta}{N_f}\right)M_0\right)^{\frac{2N_f}{N_f+1}}
m_h^{\frac{2}{N_f+1}}
+\frac{m_h^2}{8\pi}\left(1-\log m_h^2\right)
\ .
\label{VEVH}
\end{eqnarray}
The last term of (\ref{VEVH}) can be omitted because it doesn't depend
on $\theta$ and $M_0$. Therefore, the energy density $\cE$
is obtained as
\begin{eqnarray}
 \cE=-\frac{N_f+1}{8\pi}
\left(4\pi
 c\,\cos\left(\frac{\ol\theta}{N_f}\right)M_0\right)^{\frac{2N_f}{N_f+1}}
m_h^{\frac{2}{N_f+1}}\ .
\end{eqnarray}

(\ref{VEVpsibarpsi}) with the condition (\ref{muml})
corresponds to the vacuum expectation value
of the fermion mass term and therefore we conclude
\begin{eqnarray}
M_0\VEV{\ol\psi_i\psi^i}
=-\frac{N_f}{4\pi}
\left(4\pi
 c\,\cos\left(\frac{\ol\theta}{N_f}\right)M_0\right)^{\frac{2N_f}{N_f+1}}
m_h^{\frac{2}{N_f+1}} \ ,
\end{eqnarray}
which agrees with \cite{Hetrick:1995wq,Hetrick:1995yx}.

\newpage
\providecommand{\href}[2]{#2}

\bibliographystyle{utphys}

\begin{thebibliography}{99}

\bibitem{Anber:2018jdf}
  M.~M.~Anber and E.~Poppitz,
  ``Anomaly matching, (axial) Schwinger models, and high-T super
	Yang-Mills domain walls,''
  JHEP {\bf 1809} (2018) 076
  [arXiv:1807.00093 [hep-th]].

\bibitem{Coleman:1973ci}
  S.~R.~Coleman,
  ``There are no Goldstone bosons in two-dimensions,''
  Commun.\ Math.\ Phys.\  {\bf 31} (1973) 259.
  
\bibitem{Mermin:1966fe}
  N.~D.~Mermin and H.~Wagner,
  ``Absence of ferromagnetism or antiferromagnetism in one-dimensional
	or two-dimensional isotropic Heisenberg models,''
  Phys.\ Rev.\ Lett.\  {\bf 17} (1966) 1133.

\bibitem{Hohenberg:1967zz}
  P.~C.~Hohenberg,
  ``Existence of Long-Range Order in One and Two Dimensions,''
  Phys.\ Rev.\  {\bf 158} (1967) 383.

\bibitem{Anber:2018xek}
  M.~M.~Anber and E.~Poppitz,
  ``Domain walls in high-$T SU(N)$ super Yang-Mills theory and QCD(adj),''
  arXiv:1811.10642 [hep-th].

\bibitem{Unsal:2007jx}
  M.~Unsal,
  ``Magnetic bion condensation: A New mechanism of confinement and mass gap
 in four dimensions,''
  Phys.\ Rev.\ D {\bf 80} (2009) 065001
  [arXiv:0709.3269 [hep-th]].

\bibitem{Anber:2018tcj}
  M.~M.~Anber and E.~Poppitz,
  ``Two-flavor adjoint QCD,''
  Phys.\ Rev.\ D {\bf 98} (2018) no.3,  034026
  [arXiv:1805.12290 [hep-th]].

\bibitem{Yamaguchi:2018xse} 
  S.~Yamaguchi,
  ``'t Hooft anomaly matching condition and chiral symmetry breaking
 without bilinear condensate,''
  arXiv:1811.09390 [hep-th].

\bibitem{Gaiotto:2017yup}
  D.~Gaiotto, A.~Kapustin, Z.~Komargodski and N.~Seiberg,
  ``Theta, Time Reversal, and Temperature,''
  JHEP {\bf 1705} (2017) 091
  [arXiv:1703.00501 [hep-th]].

\bibitem{Sugimoto:1999tx}
  S.~Sugimoto,
  ``Anomaly cancellations in type I D-9 - anti-D-9 system and the
  USp(32) string theory,''
  Prog.\ Theor.\ Phys.\  {\bf 102} (1999) 685
  [hep-th/9905159].

\bibitem{Antoniadis:1999xk}
  I.~Antoniadis, E.~Dudas and A.~Sagnotti,
  ``Brane supersymmetry breaking,''
  Phys.\ Lett.\ B {\bf 464} (1999) 38
  [hep-th/9908023].

\bibitem{Aldazabal:1999jr}
  G.~Aldazabal and A.~M.~Uranga,
  ``Tachyon free nonsupersymmetric type IIB orientifolds via Brane -
	anti-brane systems,''
  JHEP {\bf 9910} (1999) 024
  [hep-th/9908072].

\bibitem{Angelantonj:1999ms}
  C.~Angelantonj, I.~Antoniadis, G.~D'Appollonio, E.~Dudas and A.~Sagnotti,
  ``Type I vacua with brane supersymmetry breaking,''
  Nucl.\ Phys.\ B {\bf 572} (2000) 36
  [hep-th/9911081].

\bibitem{Uranga:1999ib}
  A.~M.~Uranga,
  ``Comments on nonsupersymmetric orientifolds at strong coupling,''
  JHEP {\bf 0002} (2000) 041
  [hep-th/9912145].

\bibitem{Sugimoto:2012rt} 
  S.~Sugimoto,
  ``Confinement and Dynamical Symmetry Breaking in non-SUSY Gauge
Theory from S-duality in String Theory,''
  Prog.\ Theor.\ Phys.\  {\bf 128}, 1175 (2012)
  [arXiv:1207.2203 [hep-th]].

\bibitem{Armoni:2013ika}
  A.~Armoni,
  ``Nonsupersymmetric brane configurations, Seiberg duality,
 and dynamical symmetry breaking,''
  Phys.\ Rev.\ D {\bf 89} (2014) no.12,  125025
  [arXiv:1310.2027 [hep-th]].

\bibitem{Armoni:2014cia}
  A.~Armoni and E.~Ireson,
  ``Level-rank duality in Chern-Simons theory from a non-supersymmetric
 brane configuration,''
  Phys.\ Lett.\ B {\bf 739} (2014) 387
  [arXiv:1408.4633 [hep-th]].
               
\bibitem{Armoni:2017jkl} 
  A.~Armoni and V.~Niarchos,
  ``Phases of QCD$_3$ from Non-SUSY Seiberg Duality and Brane Dynamics,''
  Phys.\ Rev.\ D {\bf 97}, no. 10, 106001 (2018)
  [arXiv:1711.04832 [hep-th]].

\bibitem{Tanizaki:2018wtg} 
  Y.~Tanizaki,
  ``Anomaly constraint on massless QCD and the role of Skyrmions in
	chiral symmetry breaking,''
  JHEP {\bf 1808}, 171 (2018)
  [arXiv:1807.07666 [hep-th]].

\bibitem{Frishman:1992mr} 
  Y.~Frishman and J.~Sonnenschein,
  ``Bosonization and QCD in two-dimensions,''
  Phys.\ Rept.\  {\bf 223}, 309 (1993)
  [hep-th/9207017].
  
\bibitem{Komargodski:2017dmc}
  Z.~Komargodski, A.~Sharon, R.~Thorngren and X.~Zhou,
  ``Comments on Abelian Higgs Models and Persistent Order,''
  arXiv:1705.04786 [hep-th].

\bibitem{Smilga:1992hx} 
  A.~V.~Smilga,
  ``On the fermion condensate in Schwinger model,''
  Phys.\ Lett.\ B {\bf 278}, 371 (1992).

\bibitem{Witten:1983ar} 
  E.~Witten,
  ``Nonabelian Bosonization in Two-Dimensions,''
  Commun.\ Math.\ Phys.\  {\bf 92}, 455 (1984).

\bibitem{Coleman:1976uz} 
  S.~R.~Coleman,
  ``More About the Massive Schwinger Model,''
  Annals Phys.\  {\bf 101}, 239 (1976).
  
\bibitem{Hetrick:1995wq}
  J.~E.~Hetrick, Y.~Hosotani and S.~Iso,
  ``The Massive multi - flavor Schwinger model,''
  Phys.\ Lett.\ B {\bf 350} (1995) 92
  [hep-th/9502113].

\bibitem{Hetrick:1995yx}
  J.~E.~Hetrick, Y.~Hosotani and S.~Iso,
  ``The Interplay between mass, volume, vacuum angle and chiral
	condensate in N flavor QED in two-dimensions,''
  Phys.\ Rev.\ D {\bf 53} (1996) 7255
  [hep-th/9510090].

\bibitem{Coleman:1974bu}
  S.~R.~Coleman,
  ``The Quantum Sine-Gordon Equation as the Massive Thirring Model,''
  Phys.\ Rev.\ D {\bf 11} (1975) 2088.

\bibitem{Rodriguez:1996zj}
  R.~Rodriguez and Y.~Hosotani,
``Confinement and chiral condensates in 2-D QED with massive n flavor fermions,''
  Phys.\ Lett.\ B {\bf 375} (1996) 273
  [hep-th/9602029].
 
\bibitem{Coleman:1975pw} 
  S.~R.~Coleman, R.~Jackiw and L.~Susskind,
  ``Charge Shielding and Quark Confinement in the Massive Schwinger Model,''
  Annals Phys.\  {\bf 93}, 267 (1975).
  
\bibitem{Gross:1995bp}
  D.~J.~Gross, I.~R.~Klebanov, A.~V.~Matytsin and A.~V.~Smilga,
  ``Screening versus confinement in (1+1)-dimensions,''
  Nucl.\ Phys.\ B {\bf 461} (1996) 109
  [hep-th/9511104].

\bibitem{Armoni:1997ki} 
  A.~Armoni, Y.~Frishman and J.~Sonnenschein,
  ``The String tension in massive QCD in two-dimensions,''
  Phys.\ Rev.\ Lett.\  {\bf 80}, 430 (1998)
  [hep-th/9709097].

\bibitem{Armoni:1999xw}
  A.~Armoni, Y.~Frishman and J.~Sonnenschein,
  ``The String tension in two-dimensional gauge theories,''
  Int.\ J.\ Mod.\ Phys.\ A {\bf 14} (1999) 2475
  [hep-th/9903153].

\bibitem{Sugimoto:2004mh}
  S.~Sugimoto and K.~Takahashi,
  ``QED and string theory,''
  JHEP {\bf 0404} (2004) 051
  [hep-th/0403247].

\bibitem{Coleman:1973jx} 
  S.~R.~Coleman and E.~J.~Weinberg,
  ``Radiative Corrections as the Origin of Spontaneous Symmetry Breaking,''
  Phys.\ Rev.\ D {\bf 7}, 1888 (1973).

\bibitem{Green:1987sp}
  M.~B.~Green, J.~H.~Schwarz and E.~Witten,
  ``Superstring Theory. Vol. 1: Introduction,''
  Cambridge, Uk: Univ. Pr. ( 1987) 469 P. ( Cambridge Monographs On
	Mathematical Physics)


  
\end{thebibliography}

\end{document}